\documentclass[12pt,journal,onecolumn]{IEEEtran}
\usepackage{epsfig,rotating,setspace,latexsym,amsmath,epsf,amssymb,amsfonts,bm,theorem,subfigure,epstopdf}
\usepackage{cite,authblk}
\usepackage{algorithmic,algorithm}
\usepackage{setspace}
\usepackage{color}
\newtheorem{lemma}{Lemma}

\newenvironment{Proof}[1]{\medskip\par\noindent{\bf Proof:\,}\,#1}{{\mbox{\,$\blacksquare$}\par}}

\allowdisplaybreaks

\begin{document}

\title{Information Freshness in Cache Updating Systems   \thanks{This work was supported by NSF Grants CCF 17-13977 and ECCS 18-07348.}}

\author{Melih Bastopcu \qquad Sennur Ulukus\\
	\normalsize Department of Electrical and Computer Engineering\\
	\normalsize University of Maryland, College Park, MD 20742\\
	\normalsize  \emph{bastopcu@umd.edu} \qquad \emph{ulukus@umd.edu}}

\maketitle

\vspace{-1.5cm}

\begin{abstract}	
We consider a cache updating system with a source, a cache and a user. There are $n$ files. The source keeps the freshest version of the files which are updated with known rates $\lambda_i$. The cache downloads and keeps the freshest version of the files from the source with rates $c_i$. The user gets updates from the cache with rates $u_i$. When the user gets an update, it either gets a fresh update from the cache or the file at the cache becomes outdated by a file update at the source in which case the user gets an outdated update. We find an analytical expression for the average freshness of the files at the user. Next, we generalize our setting to the case where there are multiple caches in between the source and the user, and find the average freshness at the user. We provide an alternating maximization based method to find the update rates for the cache(s), $c_i$, and for the user, $u_i$, to maximize the freshness of the files at the user. We observe that for a given set of update rates for the user (resp. for the cache), the optimal rate allocation policy for the cache (resp. for the user) is a \emph{threshold policy}, where the optimal update rates for rapidly changing files at the source may be equal to zero. Finally, we consider a system where multiple users are connected to a single cache and find update rates for the cache and the users to maximize the total freshness over all users.          
\end{abstract}

\section{Introduction}
With emerging technologies such as autonomous driving, augmented reality, social networking, high-frequency automated trading, and online gaming, time sensitive information has become ever more critical. Age of information has been proposed as a performance metric to quantify the \textit{freshness} of information in communication networks. Age of information has been studied in the context of web crawling \cite{Cho03, Brewington00, Azar18, Kolobov19a}, social networks \cite{Ioannidis09}, queueing networks \cite{ Kaul12a, Costa14, Bedewy16, He16a, Kam16b, Sun17a, Najm18b, Najm17, Soysal18, Soysal19, Yates20}, caching systems \cite{Gao12, Yates17b, Kam17b, Zhong18c, Zhang18, Tang19, Yang19a, Bastopcu20c}, remote estimation \cite{Wang19a, Sun17b, Sun18b, Chakravorty18}, energy harvesting systems \cite{Bacinoglu15, Bacinoglu17, Feng18a, Feng18c, Wu18, Arafa18a, Arafa18b, Arafa18c, Arafa18f, Baknina18a, Baknina18b, Arafa17b, Arafa17a, Arafa19e, Farazi18, Leng19, Chen19}, fading wireless channels \cite{Bhat19, Ostman19}, scheduling in networks \cite{Nath17, Hsu18b, Kadota18a, Kosta17a, Bastopcu18, bastopcu_soft_updates_journal, Buyukates18c, Buyukates19b, Bastopcu20a}, multi-hop multicast networks \cite{ Zhong17a, Buyukates18, Buyukates19, Buyukates18b}, lossless and lossy source coding \cite{Zhong16, Zhong18f, Mayekar18, Ramirez19, MelihBatu1, MelihBatu2,  Bastopcu20}, computation-intensive systems \cite{Gong19, Buyukates19c, Zou19b, Arafa19a, Bastopcu19, Bastopcu20b}, vehicular, IoT and UAV systems \cite{ Elmagid18, Liu18, Elmagid19c}, reinforcement learning \cite{Ceran18, Beytur19, Elmagid19}, and so on.
 
\begin{figure}[t]
	\centering  
	\includegraphics[width=0.75\columnwidth]{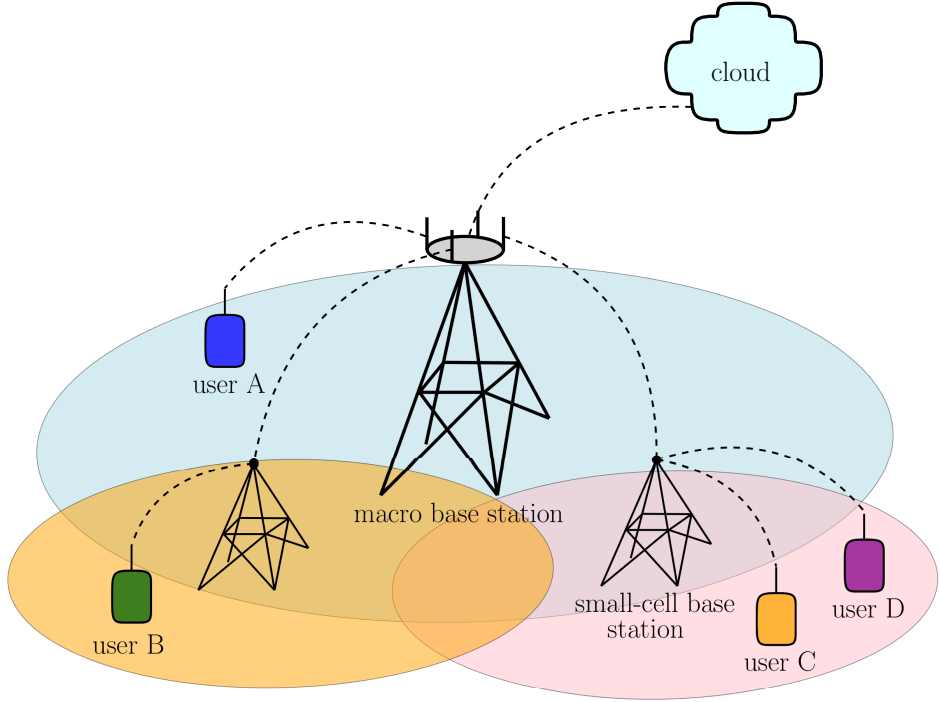}
	\caption{A cache updating system consisting of a cloud (the source), a macro base station (the first cache), a small-cell base station (the second cache), and users. The files at the source are updated with known rates. The first cache always obtains fresh files from the source. However, depending on the file status at the first cache, the second cache may not be able to obtain a fresh file all the time; the same is true for the users as well. We consider end-to-end freshness at the users.}
	\label{fig:main_model}
\end{figure} 
 
\begin{figure}[t]
	\centering  \includegraphics[width=0.75\columnwidth]{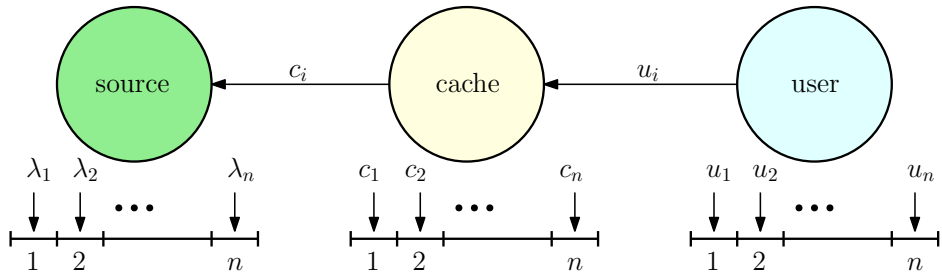}
	\caption{A cache updating system which consists of a source, a cache and a user. The $i$th file at the source is updated with rate $\lambda_i$, the cache requests updates for the $i$th file from the source with rate $c_i$, and the user requests updates for the $i$th file from the cache with rate $u_i$.}
	\label{fig:model}
\end{figure}

In this work, we consider a cache updating system that consists of a source, cache(s) and user(s). We start with the simplest system model  with a source, a single cache and a single end-user in Section~\ref{Sec:Average_freshness} (and shown in Fig.~\ref{fig:model}); generalize it to the case where there are multiple caches in between the source and the user in Section~\ref{sec:general_case} (and shown in Fig.~\ref{fig:model_gen}); and further extend it to the case where there are multiple end-users in Section~\ref{sect:mult_sol} (and shown in Fig.~\ref{fig:model_multi}). The models we study are abstractions of a real-life setting shown in Fig.~\ref{fig:main_model}. Specifically, the two-hop serial cache system considered in Section~\ref{Sec:Average_freshness} is an abstraction for the communication system between the cloud, macro base station and user $A$ in Fig.~\ref{fig:main_model}; the multi-hop serial cache system considered in Section~\ref{sec:general_case} is an abstraction for the communication system between the cloud, macro base station, small-cell base station and user $B$ in Fig.~\ref{fig:main_model} (for a three-hop system); and the multi-access caching system considered in Section~\ref{sect:mult_sol} is an abstraction for the communication system between the cloud, macro base station, small-cell base station and users $C$ and $D$ in Fig.~\ref{fig:main_model}. Example deployment scenarios for hierarchical caching systems connected wirelessly can be found in 5G-enabled vehicular networks where self-sustaining wirelessly connected caching stations are placed to enhance vehicular network capacity \cite{Zhang17}.        
 
In all these system models, the source keeps the freshest version of all the files which are updated with known rates $\lambda_i$. The cache downloads the files from the source and stores the latest downloaded versions of these files. When the cache downloads a file from the source, the file at the cache becomes fresh. After that, either the user gets the fresh file from the cache or the file at the cache becomes outdated due to a file update at the source. Thus, depending on the file status at the cache, the user may get a fresh or an outdated file. For all these system models, we derive analytical expressions for the information freshness at the end-users, and determine the updating frequencies for the intermediate caches and the end-users for maximum freshness.
 
References that are most closely related to our work are \cite{Kolobov19a, Yates17b}. Reference \cite{Kolobov19a} studies the problem of finding optimal crawl rates to keep the information in a search engine fresh while maintaining the constraints on crawl rates imposed by the websites and also the total crawl rate constraint of the search engine. Even though the freshness metric used in \cite{Kolobov19a} is similar to ours, the problem settings are different where we develop a general freshness expression for a multi-hop multi-user caching system, which differentiates our overall work from \cite{Kolobov19a}. Reference \cite{Yates17b} considers a similar model to ours where a resource constrained remote server wants to keep the items at a local cache as fresh as possible. Reference \cite{Yates17b} shows that the update rates of the files should be chosen proportional to the square roots of their popularity indices. Different from \cite{Yates17b} where the freshness of the local cache is considered, we consider the freshness at the user which is connected to the source via a single cache or multiple caches. Thus, our system model can be thought of as an extended version of the one-hop model in \cite{Yates17b}. However, our freshness metric is different than the traditional age metric used in \cite{Yates17b}, and hence, the overall work in this paper is distinct compared to \cite{Yates17b}.      

In this paper, we first consider a system where there is a source, a cache and a user (Fig.~\ref{fig:model}). We find an analytical expression for the average freshness of the files at the user. We then generalize our result to find the average freshness for the end-user when multiple caches are placed in between the source and the user (Fig.~\ref{fig:model_gen}). We impose total update rate constraints for the caches and also for the user due to limited nature of resources. Our aim is to find the update rates for the cache(s) and also for the user such that the total freshness of the files at the user is maximized. We find that the average freshness of the user is a concave function of the update rates of the caches and of the user individually, but not jointly. We provide an alternating maximization based solution where the update rates of the user (resp. of the cache) are optimized for a given set of update rates of the cache (resp. of the user). We observe that for a given set of parameters, such as update rates of the user, the optimal rate allocation policy for the other set of parameters, such as update rates at the caches, is a \emph{threshold policy}, where the files that are updated frequently at the source may not be updated by the corresponding entity. Finally, we consider a system where multiple users are connected to a single cache (Fig.~\ref{fig:model_multi}) and find update rates for the cache and for the users to maximize the total freshness over all users.         

\section{System Model, Freshness Function and Problem Formulation} \label{sect:system_model}

We consider a cache updating system where there is an information source, a cache and a user as shown in Fig.~\ref{fig:model}. The information source keeps the freshest version of $n$ files where the $i$th file is updated with exponential inter-arrival times with rate $\lambda_i$. The file updates at the source are independent of each other. A cache which is capable of storing the latest downloaded versions of all files gets the fresh files from the source. The channel between the source and the cache is assumed to be perfect and the transmission times are negligible, which is possible if the distance between the source and the cache and/or the file sizes are relatively small, as in\cite{Bacinoglu15, Bacinoglu17, Wu18, Arafa18a, Arafa18b, Arafa18c, Arafa18f,  Baknina18a, Baknina18b, Feng18a, Feng18c}. Thus, if the cache requests an update for the $i$th file, it receives the file from the source right away. The inter-update request times of the cache for the $i$th file are exponential with rate $c_i$. The cache is subject to a total update rate constraint as in \cite{Yates17b} as it is resource-constrained, i.e., $\sum_{i=1}^{n}c_i \leq C$.\footnote{We note that in practical systems, the cache's frequent update requests from the source can cause network congestion problems. Even though we impose the total update rate constraint $C$ due to the cache's limited resources, we can also imagine it to be imposed by the source in order to avoid network congestion problems.} The user requests the latest versions of the files stored in the cache. The inter-update request times of the user for the $i$th file are exponential with rate $u_i$. The channel between the user and the cache is also assumed to be perfect and the transmission times are negligible. Similarly, there is a total update rate constraint for the user, i.e., $\sum_{i=1}^{n}u_i \leq U$.

We note that each file at the source is always \textit{fresh}. However, when a file is updated at the source, the stored versions of the same file at the cache and at the user become \textit{outdated}. When the cache gets an update for an outdated file, the updated file in the cache becomes \textit{fresh} again until the next update arrival at the source. When the user requests an update for an outdated file, it might still receive an outdated version if the file at the cache is not fresh. We note that since the cache and the user are unaware of the file updates at the source, they do not know whether they have the freshest versions of the files or not. Thus, they may still request an update even though they have the freshest version of a file. 

\begin{figure}[t]
	\begin{center}
		\subfigure[]{%
			\includegraphics[width=0.49\linewidth]{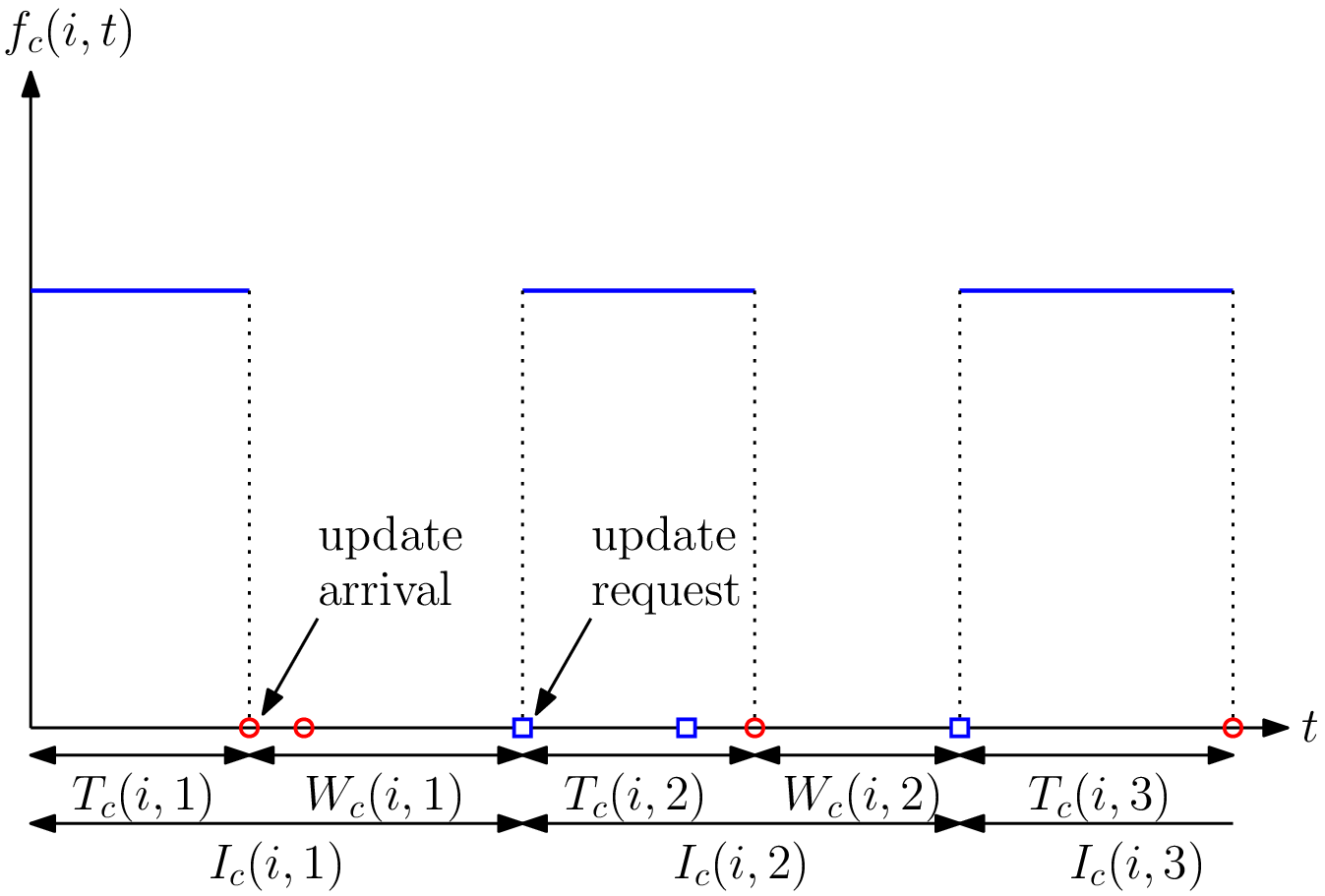}}
		\subfigure[]{%
			\includegraphics[width=0.49\linewidth]{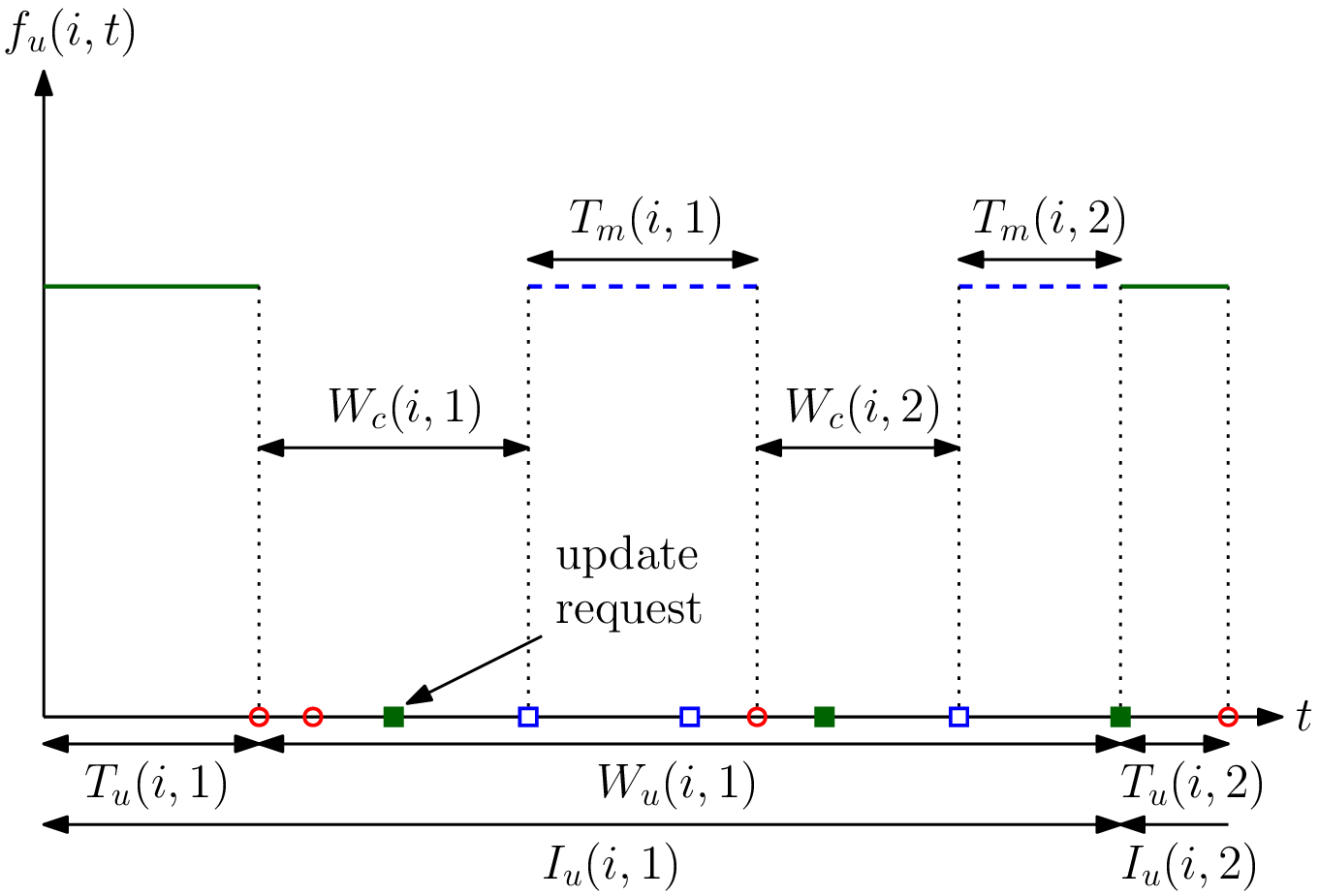}}
	\end{center}
	\caption{Sample evolution of freshness of the $i$th file (a) at the cache and (b) at the user. Red circles represent the update arrivals at the source, blue squares represent the update requests from the cache, and green filled squares represent the update requests from the user.}
	\label{Fig:age_evol}
\end{figure} 

In order to keep track of the \textit{freshness}, we define the freshness function of the $i$th file at the cache $f_{c}(i,t)$ shown in Fig.~\ref{Fig:age_evol}(a) as,
\begin{align}
f_{c}(i,t) = \begin{cases} 
1, & \text{if the $i$th file is fresh at time $t$}, \\
0, & \text{otherwise.}
\end{cases}
\end{align}
i.e., the instantaneous freshness function is a binary function taking values of fresh, ``$1$'', or not fresh, ``$0$'', at any time $t$. The binary freshness function \cite{Cho03} is different than the well-known age of information metric \cite{Kaul12a}. For example, if the information at the source does not change frequently, then the receiver may still have the freshest version of the information even though it has not received an update from the source for a while. In this case, the traditional age of information metric at the receiver would be high indicating that the information at the receiver is stale. Since the traditional age of information metric does not take into account the information change rate at the source, it may not be appropriate to measure information freshness in such systems. As the binary freshness metric compares the information at the receiver with the information stored at the source, it can measure information freshness at the receiver better than the traditional age of information in such settings.  

We denote file updates which replace an outdated file with the freshest version of the file as successful updates. We define the time interval between the $j$th and the $(j+1)$th successful updates for the $i$th file at the cache as the $j$th update cycle and denote it by $I_{c}(i,j)$. We denote the time duration when the $i$th file at the cache is fresh during the $j$th update cycle as $T_{c}(i,j)$. Then, we define the long term average freshness of the $i$th file at the cache as
\begin{align}
F_{c}(i) = \lim_{T\rightarrow\infty} \frac{1}{T} \int_{0}^{T} f_{c}(i,t)dt.    
\end{align}
Let $N$ denote the number of update cycles in the time duration $T$. Given that the system is ergodic, $\lim\limits_{T\rightarrow \infty}\frac{N}{T}$ exists and is finite. Then, similar to \cite{Cho03, Najm17, Kaul12a}, $F_{c}(i)$ is equal to
\begin{align}\label{eqn_freshness_cache}
F_{c}(i) = \lim\limits_{T\rightarrow \infty}\frac{N}{T}\left(\frac{1}{N}\sum_{j=1}^{N}T_{c}(i,j)\right) = \frac{\mathbb{E}[T_{c}(i)]}{\mathbb{E}[I_{c}(i)]}.
\end{align}

Similarly, we define $f_{u}(i,t)$ as the freshness function of the $i$th file at the user shown in Fig.~\ref{Fig:age_evol}(b). Then, the long term average freshness of the $i$th file at the user is equal to 
\begin{align}
F_{u}(i)= \frac{\mathbb{E}[T_{u}(i)]}{\mathbb{E}[I_{u}(i)]}. 
\end{align}
 Finally, we define $F_u$ as the total freshness over all files at the user as
\begin{align}\label{total_freshness_user_new}
F_u = \sum_{i=1}^{n} F_{u}(i).    
\end{align}

Our aim is to find the optimal update rates for the cache, $c_i$, and for the user, $u_i$, for $i=1,\dots,n$, such that the total average freshness of the user $F_u$ is maximized while satisfying the constraints on the total update rate for the cache, $\sum_{i=1}^{n}c_i\leq C$, and for the user, $\sum_{i=1}^{n}u_i\leq U$. Thus, our optimization problem is,
\begin{align}
\label{problem1}
\max_{\{c_i, u_i \}}  \quad &  F_u\nonumber \\
\mbox{s.t.} \quad & \sum_{i=1}^{n}c_i\leq C \nonumber \\
\quad & \sum_{i=1}^{n}u_i\leq U \nonumber \\
\quad & c_i\geq 0, \quad u_i\geq 0,\quad i=1,\dots,n.
\end{align} 

In the following section, we find analytical expressions for the long term average freshness of the $i$th file at the cache, $F_{c}(i)$, and at the user, $F_{u}(i)$, as a function of the update rate at the source $\lambda_i$, the update rate at the cache $c_i$, and the update rate at the user $u_i$. Once we find $F_u(i)$, this will determine the objective function of (\ref{problem1}) via (\ref{total_freshness_user_new}).

\section{Average Freshness Analysis for a Single Cache} \label{Sec:Average_freshness}  

In this section, we consider the system model in Fig.~\ref{fig:model}, where there is a source, a single cache and a user. First, we find an analytical expression for the long term average freshness of the $i$th file at the cache, i.e., $F_{c}(i)$ in (\ref{eqn_freshness_cache}). We note that due to the memoryless property of the exponential distribution, $T_{c}(i,j)$ which is the time duration when the $i$th file at the cache is fresh during the $j$th update cycle is exponentially distributed with parameter $\lambda_i$. Since $T_{c}(i,j)$ are independent and identically distributed (i.i.d.) over $j$, we drop index $j$, and denote a typical $T_{c}(i,j)$ as $T_{c}(i)$. Thus, we have $\mathbb{E}[T_{c}(i)] =\frac{1}{\lambda_i}$. Let $W_{c}(i,j)$ be the total duration when the $i$th file at the cache is outdated during the $j$th update cycle, i.e., $W_{c}(i,j) = I_{c}(i,j)- T_{c}(i,j)$. Note that $W_{c}(i,j)$ is also equal to the time passed until the fresh version of the $i$th file is obtained from the source after the file is outdated at the cache. We denote typical random variables for $W_{c}(i,j)$ and $I_{c}(i,j)$ by $W_{c}(i)$ and $I_{c}(i)$, respectively. As the update request times for the cache are exponentially distributed with rate $c_i$, we have $\mathbb{E}[W_{c}(i)] = \frac{1}{c_i}$. Thus, we find
\begin{align}
  \mathbb{E}[I_{c}(i)] =\mathbb{E}[T_{c}(i)]+ \mathbb{E}[W_{c}(i)] = \frac{1}{\lambda_i}+\frac{1}{c_i}.  
\end{align}
By using (\ref{eqn_freshness_cache}), we find $F_{c}(i)$ as 
\begin{align}\label{eqn_cache_freshness}
F_{c}(i) = \frac{c_i}{c_i+\lambda_i}.
\end{align}
We note that the freshness of the $i$th file at the cache $F_c(i)$ in (\ref{eqn_cache_freshness}) is an increasing function of the cache update rate $c_i$, but a decreasing function of the source update rate $\lambda_i$.\footnote{We note that \cite{Cho03} investigates several updating policies, including fixed-order updating, random-order updating, and purely random updating. The freshness metric with fixed-order updating is $F_{\text{fixed-order}} =\frac{c_i}{\lambda_i} \left(1-e^{-\frac{\lambda_i}{c_i}}\right)$, with random-order updating is $F_{\text{random-order}} = \frac{c_i}{\lambda_i}\left(1-\frac{c_i^2}{\lambda_i^2}\left(1-e^{-\frac{\lambda_i}{c_i}} \right)^2\right)$, and with purely random updating is $F_{\text{purely-random}}=\frac{c_i}{c_i+\lambda_i}$, as given in (\ref{eqn_cache_freshness}). All these three functions are monotonically decreasing in $\frac{\lambda_i}{c_i},$ and have similar forms when plotted. We adopt purely random updating in this paper due to its simplicity, and amenability to yield closed form expressions when optimized.}

Next, we find an analytical expression for the average freshness of the $i$th file at the user $F_{u}(i)$. Similar to $\mathbb{E}[T_{c}(i)]$, we have $\mathbb{E}[T_{u}(i)] = \frac{1}{\lambda_i}$ due to the memoryless property of the exponential distribution, i.e., after the user gets the fresh file, the remaining time for the next file update at the source is still exponentially distributed with rate $\lambda_i$. Similarly, we denote the time duration when the $i$th file at the user is outdated during the $j$th update cycle as $W_{u}(i,j)$ which is equal to $W_{u}(i,j) =  I_{u}(i,j)- T_{u}(i,j)$. In order for the user to get fresh updates from the cache, the cache needs to get the fresh update from the source which takes $W_{c}(i,j)$ time as discussed earlier. After the file at the cache becomes fresh, either the user gets the fresh update from the cache or the file at the source is updated, and thus the file at the cache becomes outdated again. We denote the earliest time that one of these two cases happens as $T_{m}(i)$, i.e., $T_{m}(i) = \min\{ T_{c}(i), \bar{W}_{u}(i)\}$ where $\bar{W}_{u}(i)$ is the time for the user to obtain a new update from the cache which is exponentially distributed with rate $u_i$. Thus, $T_{m}(i)$ is also exponentially distributed with rate $u_i+\lambda_i$. We note that $\mathbb{P}[T_{m}(i) = \bar{W}_{u}(i)] = \frac{u_i}{u_i+\lambda_i}$ and $\mathbb{P}[T_{m}(i) = T_{c}(i)] = \frac{\lambda_i}{u_i+\lambda_i}$. 
	 
Note that if the user gets the fresh update before the file at the cache becomes outdated which happens with probability $\mathbb{P}[T_{m}(i) = \bar{W}_{u}(i)]$, an update cycle of the $i$th file at the user is completed and thus, $f_{u}(i,t)$ is equal to $1$ again. However, if the file at the source is updated before the user gets the fresh update from the cache, then this process repeats itself, i.e., the cache initially needs to receive the fresh update which takes another $W_{c}(i,j)$ time and so on, until the user receives the fresh update from the cache. Thus, we write $ W_{u}(i,j)$ as
\begin{align}
    W_{u}(i,j) = \sum_{k = 1}^{K}W_{c}(i,k)+T_{m}(i,k), 
\end{align}
where $K$ is a geometric random variable with rate $\frac{u_i}{u_i+\lambda_i}$. Due to \cite[Prob. 9.4.1]{Yates14}, $\sum_{k = 1}^{K}W_{c}(i,k)$ and $\sum_{k = 1}^{K}T_{m}(i,k)$ are exponentially distributed with rates $ \frac{u_i c_i }{u_i+\lambda_i}$ and $u_i$, respectively. We use $W_{u}(i)$ and $I_{u}(i)$ to denote the typical random variables for $W_{u}(i,j)$ and $I_{u}(i,j)$, respectively. Thus, we have $\mathbb{E}[W_{u}(i)] = \frac{u_i+\lambda_i}{u_ic_i}+\frac{1}{u_i}$. Since $\mathbb{E}[I_{u}(i)]= \mathbb{E}[T_{u}(i)]+\mathbb{E}[W_{u}(i)]$, we get
\begin{align}
	 \mathbb{E}[I_{u}(i)] = \frac{1}{\lambda_i}+\frac{1}{u_i}+\frac{u_i+\lambda_i}{u_ic_i}. 
\end{align}
Finally, we find $F_{u}(i)$ as
\begin{align}\label{freshness_user}
	 F_{u}(i) = \frac{\mathbb{E}[T_{u}(i)]}{\mathbb{E}[I_{u}(i)]} = \frac{u_i}{u_i+\lambda_i}\frac{c_i}{c_i+\lambda_i}.
\end{align}   

We note that the freshness of the $i$th file at the user $F_{u}(i)$ in (\ref{freshness_user}) depends not only on the update rate of the user $u_i$ and file update rate at the source $\lambda_i$ but also the update rate of the cache $c_i$ as the user obtains the fresh update from the cache.\footnote{ In this paper, we assume that the channel is perfect and the transmission times are negligible. We note that $F_{u}(i)$ in (\ref{freshness_user}) can be extended to a case where the transmissions are still instantaneous but are not error free. Let $p_i$ and $q_i$ be the probabilities of successfully transmitting the $i$th file for the cache and for the user, respectively. By using [68, Prob. 9.4.1], one can show that the successful inter-update request times for the cache and for the user are exponentially distributed with rates $p_i c_i$ and $q_i u_i$, respectively. Then, the freshness of the $i$th file at the user becomes $F_{u}(i)  = \frac{q_i u_i}{q_i u_i+\lambda_i}\frac{p_i c_i}{p_i c_i+\lambda_i}.$ We observe that the freshness of the $i$th file at the user increase with $p_i$ and $q_i$ as expected.} \footnote{Even though we neglect the transmission times, as an extension to our work, in \cite{Bastopcu20c}, we consider a cache updating system where the user is able to obtain uncached files directly from the source. However, the channel between the user and the source is imperfect, and thus there is a transmission time which is exponentially distributed with rate $s_i$. In \cite{Bastopcu20c}, if the $i$th file is not cached and the cache is able to send all the file update requests from the user to the source, then the freshness of the $i$th file is equal to $F_u(i) = \frac{u_i}{u_i+\lambda_i+\frac{u_i \lambda_i}{s_i}}$. Thus, we observe that $F_u(i)$ decreases with the transmission delays as the additional term $\frac{u_i \lambda_i}{s_i}$ in the denominator decreases the freshness.} We note that $F_{u}(i)$ is an increasing function of $u_i$ and $c_i$, but a decreasing function of $\lambda_i$. We observe that $F_{u}(i)$ is an individually concave function of $u_i$ and $c_i$ but not jointly concave in $u_i$ and $c_i$, as $u_i$ and $c_i$ terms appear as a multiplication in (\ref{freshness_user}). If the user was directly connected to the source, its freshness would be equal to $ \frac{u_i}{u_i+\lambda_i}$, i.e., the first term in (\ref{freshness_user}). However, as the user is connected to the source via the cache, the freshness experienced by the user is equal to the multiplication of the freshness of the cache and the freshness of the user if the user was directly connected to the source. Note that, since $\frac{c_i}{c_i+\lambda_i}<1,$ the freshness of the user when connected to the source via a cache is smaller than the freshness it would achieve if it was directly connected to the source. 

In the following section, we find the average freshness of the caches and of the user for the general case when there are $m$ caches connected serially in between the source and the user.

\begin{figure}[t]
	\centering  \includegraphics[width=0.95\columnwidth]{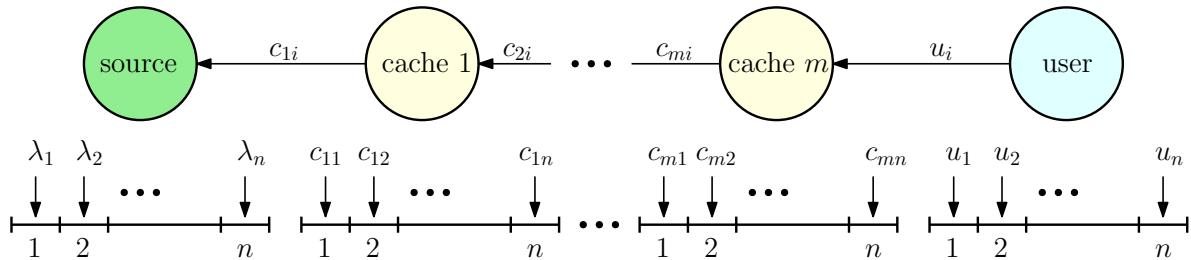}
	\caption{Generalized system model where there are $m$ serially connected caches in between the source and the user.}
	\label{fig:model_gen}
\end{figure} 

\section{Average Freshness Analysis for $M$ Caches}\label{sec:general_case}
                             
In this section, we consider a system where there are $m$ caches placed in between the source and the user, as shown in Fig.~\ref{fig:model_gen}. We denote the $r$th cache's update rate for the $i$th file as $c_{ri}$. We define $I_{c}(r,i,j)$ as the $j$th update cycle for the $i$th file at cache $r$ for $r=1,\dots,m$. Similarly, we define $T_{c}(r,i,j)$ (and $W_{c}(r,i,j)$) as the time duration when the $i$th file at cache $r$ is fresh (and outdated) during the $j$th update cycle, i.e., we have $I_{c}(r,i,j) = T_{c}(r,i,j)+W_{c}(r,i,j)$. 

Next, we find an analytical expression for the average freshness of the $i$th file at the $r$th cache $F_{c}(r,i)$ and at the user $F_{u}(i)$. In order to obtain a fresh file at cache $r$, the file at cache $r-1$ needs to be fresh for $r>1$. Similar to the derivation of $F_{u}(i)$ in (\ref{freshness_user}), after cache $r-1$ obtains the fresh file, either cache $r$ gets the fresh file from cache $r-1$ or the file at the source is updated and the file in all the caches becomes outdated. Thus, we write $W_{c}(r,i,j)$ as
\begin{align}
	 W_{c}(r,i,j) = \sum_{\ell=1}^{K_r}W_{c}(r-1,i,\ell)+T_{m}(r,i,\ell),
\end{align}
where $K_r$ is a geometric random variable with rate $\frac{c_{ri}}{c_{ri}+\lambda_i}$ and $T_{m}(r,i) = \min\{T_{c}(r,i), \bar{W}_{c}(r,i)\}$ where $T_{c}(r,i)$ and $\bar{W}_{c}(r,i)$ are exponentially distributed with rates $\lambda_i$ and $ c_{ri}$, respectively. We note that $T_{m}(r,i)$ is also exponentially distributed with rate $ c_{ri}+\lambda_i$. Then, given that $K_r= k$, we write $\mathbb{E}[W_{c}(r,i)|K_r = k]$ as
\begin{align}
	 \mathbb{E}[W_{c}(r,i)|K_r = k] = k\left(\mathbb{E}[W_{c}(r-1,i)]+\frac{1}{c_{ri}+\lambda_i}\right).
\end{align}
Then, we find $\mathbb{E}[W_{c}(r,i)] = \mathbb{E}\left[ \mathbb{E}\left[W_{c}(r,i)|K_r  \right]\right]$ as
\begin{align}
	 \mathbb{E}[W_{c}(r,i)] = \frac{c_{ri}+\lambda_i}{c_{ri}} \mathbb{E}[W_{c}(r-1,i)]+\frac{1}{c_{ri}},
\end{align}
which is equal to $\mathbb{E}[W_{c}(1,i)] = \frac{1}{c_{1i}}$ if $r=1$, and to
\begin{align}
	 \mathbb{E}[W_{c}(r,i)]  = 
	 \frac{1}{c_{ri}}+\sum_{\ell = 1}^{r-1}\frac{1}{c_{\ell i}}\prod_{p= \ell+1}^{r}\frac{c_{pi}+\lambda_i}{c_{pi}}, \quad r = 2,\dots,m.
\end{align}
Then, by using $\mathbb{E}[I_{c}(r,i)] = \mathbb{E}[T_{c}(r,i)]+\mathbb{E}[W_{c}(r,i)]$, we have
\begin{align}
	 \mathbb{E}[I_{c}(r,i)]  = \begin{cases} 
	 \frac{1}{\lambda_i}+\frac{1}{c_{1i}}, & r=1, \\
	 \left( \frac{1}{\lambda_i}+\frac{1}{c_{1i}}\right)\prod_{\ell =2}^{r}\frac{c_{\ell i}+\lambda_i}{c_{\ell i}}, & r= 2,\dots,m.
	 \end{cases}
\end{align}
Finally, we find the average freshness for the $i$th file at cache $r$ as
\begin{align}\label{avg_freshness_cache_gen}
	 F_{c}(r,i)= \frac{\mathbb{E}[T_{c}(r,i)]}{\mathbb{E}[I_{c}(r,i)]} = \prod_{\ell=1}^{r} \frac{c_{\ell i}}{c_{\ell i}+\lambda_i}, \quad r = 1,\dots,m.
\end{align}

Similarly, we find $\mathbb{E}[I_{u}(i)]$ as
\begin{align}
	 \mathbb{E}[I_{u}(i)] = \left( \frac{1}{\lambda_i}+\frac{1}{c_{1 i}}\right)\frac{u_i+\lambda_i}{u_i}\prod_{r =2}^{m}\frac{c_{r i}+\lambda_i}{c_{r i}}. 
\end{align}
Then, the average freshness of the $i$th file at the user is
\begin{align}\label{avg_freshness_user_gen}
	 F_{u}(i) = \frac{\mathbb{E}[T_{u}(i)]}{\mathbb{E}[I_{u}(i)] } = \frac{u_i}{u_i+\lambda_i} \prod_{r=1}^{m} \frac{c_{r i}}{c_{r i}+\lambda_i}.
\end{align}
	 
Thus, we observe from (\ref{avg_freshness_cache_gen}) that, for the general system, the average freshness experienced by cache $r$ for $r>1$  is equal to the multiplication of the freshness of cache $r-1$ with the freshness of cache $r$ when cache $r$ is directly connected to the source. We observe from (\ref{avg_freshness_user_gen}) that the same structure is valid for the freshness of the user as well. We also note that the average freshness expression in (\ref{avg_freshness_user_gen}) reduces to the expression in (\ref{freshness_user}) found in Section~\ref{Sec:Average_freshness}, when $m=1$. Finally, as an explicit example of the expression in (\ref{avg_freshness_user_gen}), if we have $m=2$ caches between the source and the user, the freshness at the user is
\begin{align}
    F_{u}(i) = \frac{u_i}{u_i+\lambda_i}\frac{c_{1i}}{c_{1i}+\lambda_i}\frac{c_{2i}}{c_{2i}+\lambda_i}.
\end{align}

In the following section, we solve the optimization problem in (\ref{problem1}) for the system with a single cache by using the freshness expression $F_u(i)$ found in (\ref{freshness_user}) in Section~\ref{Sec:Average_freshness}.  

\section{Freshness Maximization for a System with a Single Cache} \label{sect:opt_soln}

In this section, we consider the optimization problem in (\ref{problem1}) for a system with a single cache. Using $F_u(i)$ in (\ref{freshness_user}) and $F_u$ in (\ref{total_freshness_user_new}), we rewrite the freshness maximization problem as   
\begin{align}
\label{problem1_mod}
\max_{\{c_i, u_i \}}  \quad &  \sum_{i=1}^{n}\frac{u_i}{u_i+\lambda_i}\frac{c_i}{c_i+\lambda_i}\nonumber \\
\mbox{s.t.} \quad & \sum_{i=1}^{n}c_i\leq C \nonumber \\
\quad & \sum_{i=1}^{n}u_i\leq U \nonumber \\
\quad & c_i\geq 0,\quad u_i\geq 0,\quad i=1,\dots,n.
\end{align}
We introduce the Lagrangian function \cite{Boyd04} for (\ref{problem1_mod}) as 
\begin{align}
\mathcal{L} =-\sum_{i=1}^{n}\frac{u_i}{u_i+\lambda_i}\frac{c_i}{c_i+\lambda_i} +\beta\left(  \sum_{i=1}^{n}c_i- C\right)+\theta\left(\sum_{i=1}^{n}u_i- U  \right)- \sum_{i=1}^{n}\nu_i c_i -\sum_{i=1}^{n}\eta_i u_i,
\end{align}
where $\beta\geq 0$, $\theta\geq 0$, $\nu_i\geq0$ and $\eta_i\geq 0 $. Then, we write the KKT conditions as 
\begin{align}
\frac{\partial \mathcal{L}}{\partial c_i} &= -\frac{u_i}{u_i+\lambda_i}\frac{\lambda_i}{\left(c_i+\lambda_i\right)^2} +\beta-\nu_i = 0, \label{KKT1}\\
\frac{\partial \mathcal{L}}{\partial u_i} &= -\frac{c_i}{c_i+\lambda_i}\frac{\lambda_i}{\left(u_i+\lambda_i\right)^2} +\theta-\eta_i = 0,\label{KKT2}
\end{align}
for all $i$. The complementary slackness conditions are
\begin{align}
\beta\left(  \sum_{i=1}^{n}c_i- C\right) &= 0, \label{CS1}\\ 
\theta\left(\sum_{i=1}^{n}u_i- U  \right) &= 0, \label{CS2}\\ 
\nu_i c_i &= 0,\label{CS3} \\
\eta_i u_i &= 0. \label{CS4}
\end{align}

The objective function in (\ref{problem1_mod}) is not jointly concave in $c_i$ and $u_i$ since $c_i$s and $u_i$s appear as multiplicative terms in the objective function. However, for given $c_i$s, the objective function in (\ref{problem1_mod}) is concave in $u_i$. Similarly, for given $u_i$s, the objective function in (\ref{problem1_mod}) is concave in $c_i$. Thus, we apply an alternating maximization based method \cite{bertsekas, AlterMin, iterminimization2,Bastopcu20} to find $(c_i, u_i)$ pairs such that (\ref{KKT1}) and (\ref{KKT2}) are satisfied for all $i$.\footnote{The proposed alternating maximization based method finds $(c_i,u_i)$ pairs that satisfy the first order optimality conditions, i.e., the KKT conditions in (\ref{KKT1})-(\ref{CS4}). We note that as the optimization problem in (\ref{problem1_mod}) is not a convex optimization problem, the solutions obtained from the alternating maximization based method are not globally optimal. Similarly, solutions obtained for the optimization problems in (\ref{problem2}) and (\ref{problem3}) are locally optimal as well.} 

Starting with initial $u_i$s, we find the optimum update rates for the cache, $c_i$s, such that the total update rate constraint for the cache, i.e., $\sum_{i=1}^{n}c_i \leq C$, and the feasibility of the update rates, i.e., $c_i\geq0 $ for all $i$, are satisfied. Then, for given $c_i$s, we find the optimum update rates for the user, $u_i$s, such that the total update rate constraint for the user, i.e., $\sum_{i=1}^{n}u_i \leq U$, and the feasibility of the update rates, i.e., $u_i\geq0 $ for all $i$, are satisfied. We repeat these steps until the KKT conditions in (\ref{KKT1}) and (\ref{KKT2}) are satisfied.

For given $u_i$s with $u_i>0$, we rewrite (\ref{KKT1}) as 
\begin{align}
(c_i+\lambda_i)^2 = \frac{1}{\beta-\nu_i}\frac{u_i\lambda_i}{u_i+\lambda_i}
\end{align}
Then, we find $c_i$ as 
\begin{align}\label{soln_c_i}
c_i = \frac{1}{\sqrt{\beta-\nu_i}}\sqrt{\frac{u_i \lambda_i}{u_i+\lambda_i}} -\lambda_i,
\end{align}                                 
for all $i$ with $u_i>0$. If $c_i>0$, we have $ \nu_i = 0$ from (\ref{CS3}). Thus, we have
\begin{align}\label{soln_c_i_2}
c_i = \left(\frac{1}{\sqrt{\beta}} \sqrt{\frac{u_i \lambda_i}{u_i+\lambda_i}}-\lambda_i\right)^+,
\end{align}
for all $i$ with $u_i>0$, where $(x)^+ = \max(x,0)$. Note that $c_i>0$ requires $ \frac{1}{\lambda_i}\frac{u_i}{u_i+\lambda_i}>\beta$ which also implies that if $ \frac{1}{\lambda_i}\frac{u_i}{u_i+\lambda_i}\leq \beta$, then we must have $c_i=0$. Thus, for given $u_i$s, we see that the optimal rate allocation policy for the cache is a \emph{threshold policy} in which the optimal update rates are equal to zero when the files are updated too frequently at the source, i.e., when the corresponding $\lambda_i$s are too large.\footnote{As the stored versions of the files that change too frequently at the source become obsolete too quickly at the user, the freshness of these files at the user will be small. That is why, instead of updating these files, with the threshold policy, the files that change less frequently at the source are updated more which brings higher contribution to the overall freshness at the user.} \footnote{ As a result of the threshold policy, the user may not receive the fresh versions of the files that change too frequently at the source which can be undesirable especially if obtaining the fresh versions of some files is more important than the others. In order to address this problem, we can introduce an importance factor for each file $\mu_i$. Then, we can rewrite the overall freshness at the user $F_u$ in (\ref{total_freshness_user_new}) as $F_u = \sum_{i=1}^{n} \mu_i F_u(i)$. We note that by solving the optimization problem in (\ref{problem1}) with a freshness expression with importance factors, the important files that change too frequently at the source may be updated by the cache and also by the user.}

Next, we solve for $u_i$s for given $c_i$s with $c_i>0$. We rewrite (\ref{KKT2}) as
\begin{align}
(u_i+\lambda_i)^2=\frac{1}{\theta-\eta_i} \frac{c_i\lambda_i}{c_i+\lambda_i}.
\end{align}
Then, we find $u_i$ as 
\begin{align}
u_i =  \frac{1}{\sqrt{\theta-\eta_i}}\sqrt{\frac{c_i \lambda_i}{c_i+\lambda_i}}-\lambda_i,
\end{align}
for all $i$ with $c_i>0$. If $u_i >0$, we have $\eta_i= 0$ from (\ref{CS4}). Thus, we have
\begin{align}
u_i = \left( \frac{1}{\sqrt{\theta}}\sqrt{\frac{c_i\lambda_i}{c_i+\lambda_i}}-\lambda_i\right)^+.
\end{align} 
Similarly, $u_i >0$ requires $\frac{1}{\lambda_i}\frac{c_i}{c_i+\lambda_i}> \theta$ which implies that if $ \frac{1}{\lambda_i}\frac{c_i}{c_i+\lambda_i}\leq \theta$, then we must have $u_i = 0$. Thus, for given $c_i$s, we see that the optimal rate allocation policy for the user is also a threshold policy in which the optimal update rates are equal to zero when the files are updated too frequently at the source, i.e., when the corresponding $\lambda_i$s are too large. 

In the following lemma, we show that if the update rate of the cache $c_i$ (resp. of the user $u_i$) is equal to zero for the $i$th file, then the update rate of the user $u_i$ (resp. of the cache $c_i$) for the same file must also be equal to zero. 

\begin{lemma}\label{lemma1}
	In the optimal policy, if $c_i=0$, then we must have $u_i = 0$; and vice versa.
\end{lemma}

\begin{Proof}
	Assume for contradiction that in the optimal policy, there exist update rates with $c_i = 0$ and $u_i>0$. We note that average freshness of this file at the user is equal to zero, i.e., $F_{u}(i) = 0,$ as $c_i = 0$. We can increase the total freshness of the user $F_u$ by decreasing $u_i$ to zero and increasing one of $u_j$s with $c_j>0$. Thus, we reach a contradiction and in the optimal policy, if $c_i = 0$, we must have $u_i = 0$. For the update rates with $c_i > 0$ and $u_i=0$, one can similarly show that  if $u_i = 0$, then we must have $c_i =0$. 
\end{Proof}

In the following lemma, we show that the total update rate constraints for the cache, i.e., $ \sum_{i=1}^{n}c_i\leq C$, and for the user, i.e., $ \sum_{i=1}^{n}u_i \leq U$, must be satisfied with equality.

\begin{lemma}\label{lemma2}
	In the optimal policy, we must have $ \sum_{i=1}^{n}c_i= C$ and $ \sum_{i=1}^{n}u_i = U$.
\end{lemma}

\begin{Proof}
	 Assume for contradiction that in the optimal policy, we have  $\sum_{i=1}^{n}c_i< C$. As the objective function in (\ref{problem1_mod}) is an increasing function of $c_i$, we can increase the total freshness of the user $F_u$ by increasing one of $c_i$ with $u_i >0$ until the total update rate constraint for the cache is satisfied with equality, i.e., $\sum_{i=1}^{n}c_i= C$. Thus, we reach a contradiction and in the optimal policy, we must have $\sum_{i=1}^{n}c_i= C$. By using a similar argument, we can also show that in the optimal policy, we must have $\sum_{i=1}^{n}u_i = U$.    
\end{Proof}

In the following lemma, we identify a property of the optimal cache update rates $c_i$ for given user update rates $u_i$. To that end, for given $u_i$s, let us define $\phi_i$s as
\begin{align} \label{phi-i-defn}
\phi_i = \frac{1}{\lambda_i}\frac{u_i}{u_i+\lambda_i}.
\end{align}
This lemma will be useful for solving for $c_i$ given $u_i$ using (\ref{soln_c_i_2}).

\begin{lemma}\label{lemma3}
	For given $u_i$s, if $c_i>0$ for some $i$, then we have $c_j> 0$ for all $j$ with $\phi_j\geq \phi_i$.
\end{lemma}

\begin{Proof}
    As we noted earlier, from (\ref{soln_c_i_2}), $c_i>0$ implies $\phi_i>\beta$. Thus, if $\phi_j\geq \phi_i$, then we have $\phi_j>\beta$, which further implies $c_j>0$.
\end{Proof}

\begin{figure}[t]
	\centering  \includegraphics[width=0.6\columnwidth]{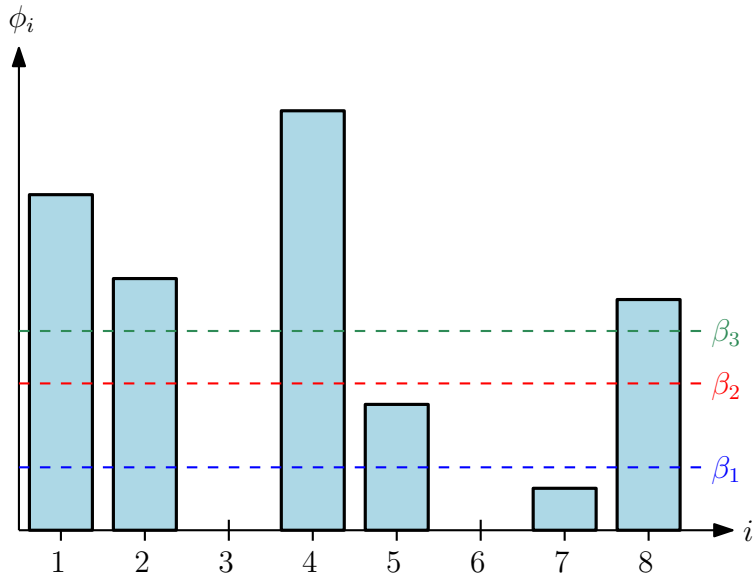}
	\caption{For given $u_i$s, we show $\phi_i$s calculated in (\ref{phi-i-defn}) for $n=8$.}
	\label{fig:upt_pol}
\end{figure}

Next, we describe the overall solution for the single cache setting. We start with a set of initial $u_i$s. We obtain $\phi_i$ from $u_i$ using (\ref{phi-i-defn}). We will utilize Fig.~\ref{fig:upt_pol} to describe the steps of the solution visually. We plot $\phi_i$ in Fig.~\ref{fig:upt_pol}. Note that if $u_i=0$ then $\phi_i=0$, and vice versa. First, we choose $c_i = 0$ for the files with $u_i = 0$ due to Lemma~\ref{lemma1}, i.e., in Fig.~\ref{fig:upt_pol}, we choose $c_3$ and $c_6$ as zero. Next, we find the remaining $c_i$s with $u_i>0$. For that, we rewrite (\ref{soln_c_i_2}) as
\begin{align}\label{policy_finder}
    c_i = \frac{\lambda_i}{\sqrt{\beta}}\left(\sqrt{\phi_i} -\sqrt{\beta}\right)^+
\end{align}
Due to Lemma~\ref{lemma2}, in the optimal policy, we must have $\sum_{i=1}^{n}c_i = C$. Assuming that $\phi_i\geq\beta$ for all $i$, i.e., by ignoring $(\cdot)^+$ in (\ref{soln_c_i_2}) and (\ref{policy_finder}), we solve $\sum_{i=1}^{n}c_i = C$ for $\beta$. Then, we compare the smallest $\phi_i$ with $\beta$. If the smallest $\phi_i$ is larger than or equal to $\beta$, it implies that $c_i>0$ for all $i$ due to Lemma~\ref{lemma3}, and we have obtained the optimal $c_i$ values for given $u_i$s. If the smallest $\phi_i$ is smaller than $\beta$, it implies that the corresponding $c_i$ was negative and it must be chosen as zero. In this case, we choose $c_i=0$ for the smallest $\phi_i$. In the example in Fig.~\ref{fig:upt_pol}, if the $\beta$ we found is $\beta_1$, then since $\phi_7< \beta_1$ we choose $c_7$ as zero. Then, we repeat this process again until the smallest $\phi_i$ among the remaining $c_i$s satisfies $\phi_i\geq \beta$. For example, in Fig.~\ref{fig:upt_pol}, the next $\beta$ found by using only indices $1, 2, 4, 5, 8$ may be $\beta_2$. Since $\phi_5< \beta_2$, we choose $c_5 = 0$. In the next iteration, the $\beta$ found by using indices $1, 2, 4, 8$ may be $\beta_3$. Since $\phi_8>\beta_3$, we stop the process and find $c_i$ for $i= 1,2,4,8$ from (\ref{soln_c_i_2}) or (\ref{policy_finder}) by using $\beta_3$ in Fig.~\ref{fig:upt_pol}. This concludes finding $c_i$s for given $u_i$s. Next, for given $c_i$s, we find $u_i$s by following a similar procedure. We keep solving for $c_i$s for given $u_i$s, and $u_i$s for given $c_i$s, until $(c_i,u_i)$ pairs converge. 

In the following section, we provide a solution for the general system with multiple caches. 

\section{Freshness Maximization for the General System}\label{sect:gen_soln}

In this section, we provide a solution for the general system shown in Fig.~\ref{fig:model_gen}, where there are $m$ caches in between the source and the user. We define $C_r$ as the total update rate of cache $r$, i.e., $\sum_{i=1}^{n}c_{ri}\leq C_r$. Using $F_{u}(i)$ in (\ref{avg_freshness_user_gen}), we rewrite the optimization problem in (\ref{problem1}) as
\begin{align}
\label{problem2}
\max_{\{c_{ri}, u_i \}}  \quad &  \sum_{i=1}^{n}\frac{u_i}{u_i+\lambda_i} \prod_{r=1}^{m} \frac{c_{r i}}{c_{r i}+\lambda_i}\nonumber \\
\mbox{s.t.} \quad & \sum_{i=1}^{n}c_{ri}\leq C_r, \quad r={1,\dots,m} \nonumber \\
\quad & \sum_{i=1}^{n}u_i\leq U \nonumber \\
\quad & c_{ri}\geq 0,\quad u_i\geq 0,\quad r={1,\dots,m},\quad i=1,\dots,n.
\end{align}
We introduce the Lagrangian function for (\ref{problem2}) as 
\begin{align}
\mathcal{L} =&-\sum_{i=1}^{n}\frac{u_i}{u_i+\lambda_i} \prod_{r=1}^{m} \frac{c_{r i}}{c_{r i}+\lambda_i} +\sum_{r=1}^{m}\beta_r\left(  \sum_{i=1}^{n}c_{r i}- C_r\right)+\theta\left(\sum_{i=1}^{n}u_i- U  \right)\nonumber\\
&- \sum_{r=1}^{m}\sum_{i=1}^{n}\nu_{r i} c_{r i} -\sum_{i=1}^{n}\eta_i u_i,
\end{align}
where $\beta_r\geq 0$, $\theta\geq 0$, $\nu_{r i}\geq0$ and $\eta_i\geq 0$. Then, we write the KKT conditions as 
\begin{align}
\frac{\partial \mathcal{L}}{\partial c_{ri}} &= -\frac{u_i}{u_i+\lambda_i}\prod_{\ell\neq r} \frac{c_{\ell i}}{c_{\ell i}+\lambda_i}\frac{\lambda_i}{\left(c_{ri}+\lambda_i\right)^2} +\beta_r-\nu_{ri} = 0, \label{KKT2_1}\\
\frac{\partial \mathcal{L}}{\partial u_i} &= -\frac{\lambda_i}{\left(u_i+\lambda_i\right)^2}\prod_{r=1}^{m} \frac{c_{r i}}{c_{r i}+\lambda_i} +\theta-\eta_i = 0,\label{KKT2_2}
\end{align}
for all $r$ and $i$. The complementary slackness conditions are
\begin{align}
\beta_r\left(  \sum_{i=1}^{n}c_{r i}- C_r\right) &= 0, \label{CS2_1}\\ 
\theta\left(\sum_{i=1}^{n}u_i- U  \right) &= 0, \label{CS2_2}\\ 
\nu_{ri} c_{ri} &= 0, \label{CS2_3} \\
\eta_i u_i &= 0. \label{CS2_4}
\end{align}

The objective function in (\ref{problem2}) is not jointly concave in $c_{ri}$ and $u_i$. However, for given $c_{ri}$s, the objective function in (\ref{problem2}) is concave in $u_i$, and for a given $u_i$ and $c_{\ell i}$s for all $\ell \neq r$, the objective function in (\ref{problem2}) is concave in $c_{r i}$. Thus, similar to the solution approach in Section~\ref{sect:opt_soln}, we apply an alternating maximization based method to find $(c_{1 i},\dots, c_{r i}, u_i)$ tuples such that (\ref{KKT2_1}) and (\ref{KKT2_2}) are satisfied for all $r$ and $i$. 

Starting with initial $u_i$ and $c_{\ell i}$s for $\ell \neq r$, we find the optimum update rates for cache $r$, $c_{ri}$s, such that the total update rate constraint for the cache, i.e., $\sum_{i=1}^{n}c_{ri} \leq C_r$, and the feasibility of the update rates, i.e., $c_{ri}\geq0 $ for all $i$, are satisfied. We repeat this step for all $r$. Then, for given $c_{ri}$s, we find the optimum update rates for the user, $u_i$s, such that the total update rate constraint for the user, i.e., $\sum_{i=1}^{n}u_i \leq U$, and the feasibility of the update rates, i.e., $u_i\geq0 $ for all $i$, are satisfied. We repeat these steps until the KKT conditions in (\ref{KKT2_1}) and (\ref{KKT2_2}) are satisfied.

For given $u_i$s with $u_i>0$, and $c_{\ell i}$ with $c_{\ell i}>0,$ for $\ell\neq r$, we rewrite (\ref{KKT2_1}) as 
\begin{align}
(c_{r i}+\lambda_i)^2 = \frac{\sigma_i\lambda_i}{\beta_r-\nu_{r i}},
\end{align}
where $\sigma_i= \frac{u_i}{u_i+\lambda_i} \prod_{\ell\neq r} \frac{c_{\ell i}}{c_{\ell i}+\lambda_i}$. Then, we find $c_{r i}$ as 
\begin{align}\label{soln_gen_c_i}
c_{r i} =  \sqrt{\frac{\sigma_i \lambda_i}{\beta_r-\nu_{r i}}}-\lambda_i ,
\end{align}                                 
for all $i$ with $u_i>0$ and $c_{\ell i}>0$ for $\ell\neq r$. If $c_{ri}>0$, we have $ \nu_{ri} = 0$ from (\ref{CS2_3}). Thus, we have
\begin{align}\label{soln_gen_c_i_2}
c_{ri} = \left( \sqrt{\frac{\sigma_i \lambda_i}{\beta_r}}-\lambda_i\right)^+.
\end{align}
Note that $c_{ri}>0$ requires $\frac{\sigma_i}{\lambda_i}>\beta_r$, i.e., $\frac{1}{\lambda_i}\frac{u_i}{u_i+\lambda_i} \prod_{\ell\neq r} \frac{c_{\ell i}}{c_{\ell i}+\lambda_i}>\beta_r$ which also implies that if $ \frac{1}{\lambda_i}\frac{u_i}{u_i+\lambda_i} \prod_{\ell\neq r} \frac{c_{\ell i}}{c_{\ell i}+\lambda_i}\leq \beta_r$, then we must have $c_{ri}=0$. We repeat this step for $r = 1,\dots, m$.      

Next, we solve for $u_i$s for given $c_{r i}$s for all $r$ with $c_{ri}>0$. We rewrite (\ref{KKT2_2}) as
\begin{align}
(u_i+\lambda_i)^2 = \frac{\rho_i\lambda_i}{\theta-\eta_i},
\end{align}
where $\rho_i =\prod_{r=1}^{m} \frac{c_{r i}}{c_{r i}+\lambda_i} $. Then, we find $u_i$ as 
\begin{align}
u_i =  \sqrt{\frac{\rho_i\lambda_i}{\theta-\eta_i}}-\lambda_i,
\end{align}
for all $i$ with $c_{r i}>0$ for all $r$. If $u_i >0$, we have $\eta_i= 0$ from (\ref{CS2_4}). Thus, we have
\begin{align}\label{opt_rate_user_gen}
u_i = \left( \sqrt{\frac{\rho_i \lambda_i}{\theta}}-\lambda_i\right)^+.
\end{align} 
Similarly, $u_i >0$ requires $\frac{\rho_i}{\lambda_i}>\theta$, i.e.,
$\frac{1}{\lambda_i}\prod_{r=1}^{m}\frac{c_{r i}}{c_{r i}+\lambda_i}> \theta$ which implies that if $ \frac{1}{\lambda_i}\prod_{r=1}^{m}\frac{c_{r i}}{c_{r i}+\lambda_i}\leq \theta$, then we must have $u_i = 0$. Thus, similar to the results in Section \ref{sect:opt_soln}, for given $c_{r i}$s (resp. for given $u_i$s and $c_{\ell i}$s for all $\ell\neq r$), we observe that the optimal rate allocation policy for the user (resp. for cache $r$) is a threshold policy and the update rates for the user (resp. for cache $r$) are equal to zero for the files with very large $\lambda_i$s.

Similar to Lemma \ref{lemma1}, one can show that in the optimal policy, if $c_{ri} = 0$ for some $r$, then we must have $c_{\ell i} = 0$ for all $\ell\neq r$ and $u_i =0$. Furthermore, if $u_i = 0$ for some $i$, then we must have $c_{r i} = 0$ for all $r$. One can also show that the total update rate constraints for cache $r$, i.e., $ \sum_{i=1}^{n}c_{r i}\leq C_r$, for all $r$ and for the user, i.e., $ \sum_{i=1}^{n}u_i \leq U$, should be satisfied with equality as the objective function in (\ref{problem2}) is an increasing function of $c_{r i}$ and $u_i$. Thus, in the optimal policy, we must have $ \sum_{i=1}^{n}c_{r i}= C_r$ for all $r$ and $ \sum_{i=1}^{n}u_i = U$.

We note from (\ref{soln_gen_c_i_2}) that the update rates of cache $r$ directly depend on the update rates of the other caches as well as the update rates of the user. Similarly, we note from (\ref{opt_rate_user_gen}) that the update rates of the user directly depend on the update rates of all caches. In order to find the overall solution, for given initial $u_i$s and $c_{\ell i}$ for all $\ell\neq r$, we choose $c_{r i} = 0$ for the files with $ u_i = 0$ or $c_{\ell i} = 0$ for some $\ell$. We solve $\sum_{i=1}^{n}c_{r i} = C_r$ for $\beta_r$ and then, find $c_{r i}$s by using (\ref{soln_gen_c_i_2}) similar to the solution method in Section \ref{sect:opt_soln}. We repeat this step for $r= 1,\dots,m$. Next, for given $c_{r i}$s, we find $u_i$s by following a similar procedure. We keep updating these parameters until $(c_{1 i},\dots,c_{r i},u_i)$ tuples converge.

In the following section, we find rate allocations for a system with a source, a single cache, and multiple users.

\section{Freshness Maximization for a System with Multiple Users}\label{sect:mult_sol}

\begin{figure}[t]
	\centering  \includegraphics[width=0.75\columnwidth]{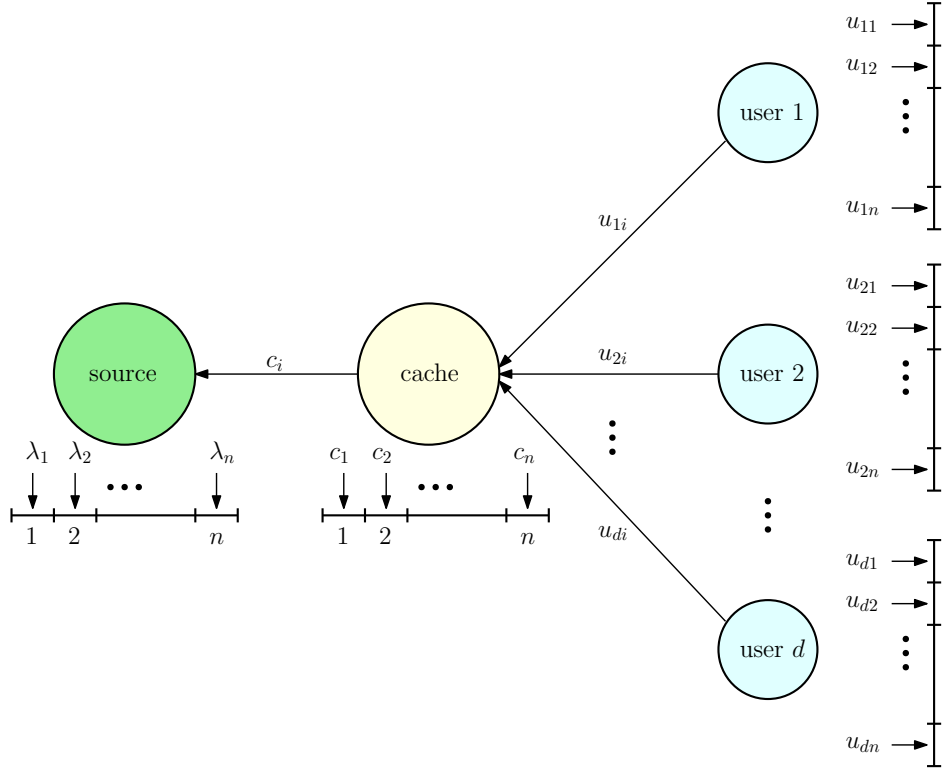}
	\caption{A cache updating system with a source, a single cache and $d$ users.}
	\label{fig:model_multi}
\end{figure}

In this section, we consider a system where there is a source, a single cache and $d$ users connected to the cache, as shown in Fig.~\ref{fig:model_multi}. Our aim is to find the update rates for the cache and for the users such that the overall freshness experienced by the users is maximized. 

Let the $k$th user's update rate for the $i$th file be $u_{ki}$. Each user is subject to a total update rate constraint as $\sum_{i=1}^n u_{ki}\leq U_k$, for $k = 1,\dots, d,$ and the cache is subject to a total update rate constraint as $\sum_{i=1}^n c_{i}\leq C$. From Section~\ref{Sec:Average_freshness}, the average freshness of the $i$th file at the $k$th user is $F_u(k,i) = \sum_{i=1}^{n} \frac{u_{ki}}{u_{ki}+\lambda_i}\frac{c_i}{c_i+\lambda_i}$. Then, we write the freshness maximization problem as
\begin{align}
\label{problem3}
\max_{\{c_i, u_{ki} \}}  \quad & \sum_{k=1}^{d}\sum_{i=1}^{n} \frac{u_{ki}}{u_{ki}+\lambda_i}\frac{c_i}{c_i+\lambda_i}\nonumber \\
\mbox{s.t.} \quad & \sum_{i=1}^{n}c_i\leq C \nonumber \\
\quad & \sum_{i=1}^{n}u_{ki}\leq U_k, \quad k= 1,\dots,d \nonumber \\
\quad & c_i\geq 0,\quad u_{ki}\geq 0, \quad k = 1,\dots,d, \quad i=1,\dots,n.
\end{align}
We introduce the Lagrangian function for (\ref{problem3}) as
\begin{align}
\mathcal{L} =&  -\sum_{k=1}^{d}\sum_{i=1}^{n} \frac{u_{ki}}{u_{ki}+\lambda_i}\frac{c_i}{c_i+\lambda_i}+\beta\left( \sum_{i=1}^{n}c_i-C\right)+\sum_{k=1}^{d}\theta_k\left( \sum_{i=1}^{n}u_{ki}- U_k\right) -\sum_{i=1}^{n}\nu_ic_i \nonumber\\&-\sum_{k=1}^{d}\sum_{i=1}^{n}\eta_{ki}u_{ki}, 
\end{align}
where $\beta \geq 0$, $\theta_k\geq 0$, $\nu_{i}\geq0$ and $\eta_{ki}\geq 0$.  We write the KKT conditions as 
\begin{align}
\frac{\partial \mathcal{L}}{\partial c_i}&= -\frac{\lambda_i}{(c_i+\lambda_i)^2}\sum_{k=1}^{d} \frac{u_{ki}}{u_{ki}+\lambda_i}+\beta-\nu_i = 0,\label{KKT3_1} \\
\frac{\partial \mathcal{L}}{\partial u_{ki}}&= -\frac{\lambda_i}{(u_{ki}+\lambda_i)^2} \frac{c_{i}}{c_{i}+\lambda_i}+\theta_k-\eta_{ki} = 0,\label{KKT3_2} 
\end{align} 
for all $k$ and $i$. The complementary slackness conditions are 
\begin{align}
\beta\left( \sum_{i=1}^{n}c_i- C\right) &= 0, \label{CS3_1}\\
\theta_k\left( \sum_{i=1}^{n}u_{ki}- U_k\right) &= 0, \label{CS3_2}\\
\nu_ic_i &= 0,\label{CS3_3}\\
\eta_{ki}u_{ki} &= 0.\label{CS3_4}
\end{align} 

The objective function in (\ref{problem3}) is not jointly concave in $c_{i}$ and $u_{ki}$. However, for given $u_{ki}$s, the objective function in (\ref{problem3}) is concave in $c_i$, and for given $c_i$s, the objective function in (\ref{problem3}) is jointly concave in all $u_{ki}$. Thus, similar to the solution approach used in Section \ref{sect:opt_soln}, we apply an alternating maximization based method to find $(c_{i}, u_{ki})$ such that (\ref{KKT3_1}) and (\ref{KKT3_2}) are satisfied for all $k$ and $i$. 

For given $u_{ki}$s, we find $c_i$ as
\begin{align}\label{c_i_soln_problem3}
c_i = \frac{1}{\sqrt{\beta-\nu_i}}\sqrt{\sum_{k=1}^{d} \frac{u_{ki}\lambda_i}{u_{ki}+\lambda_i}}-\lambda_i,
\end{align}
for all $i$. If $c_i>0$, we have $\nu_i = 0$ due to (\ref{CS3_3}). 
Thus, we have
\begin{align}\label{c_i_soln3}
c_i = \left( \frac{1}{\sqrt{\beta}}\sqrt{\sum_{k=1}^{d} \frac{u_{ki}\lambda_i}{u_{ki}+\lambda_i}}-\lambda_i\right)^+.
\end{align}
Similarly, the optimal rate allocation policy is a threshold policy where if $\frac{1}{\lambda_i}\left( \sum_{k=1}^{d} \frac{u_{ki}}{u_{ki}+\lambda_i} \right) <\beta$, then we have $c_i=0$. 

Next, for a given $c_i$ with $c_i>0$, we find $u_{ki}$ as 
\begin{align}\label{u_ki_soln_problem3}
u_{ki} =  \frac{1}{\sqrt{\theta_k-\eta_{ki}}}\sqrt{ \frac{c_{i}\lambda_i}{c_{i}+\lambda_i}}-\lambda_i,
\end{align}
for all $i$. If $u_{ki}>0$, we have $\eta_{ki} = 0$ due to (\ref{CS3_4}). Thus, we have
\begin{align}\label{u_ki_soln3}
u_{ki} = \left( \frac{1}{\sqrt{\theta_k}}\sqrt{ \frac{c_{i}\lambda_i}{c_{i}+\lambda_i}}-\lambda_i\right)^+.
\end{align}
Thus, we note that the optimal rate allocation policy is a threshold policy where if $\frac{1}{\lambda_i} \frac{c_{i}}{c_{i}+\lambda_i} <\theta_k$, then we have $u_{ki}=0$. We observe that the optimal rates for the users depend directly on $\lambda_i$ and $c_i$. As the update rates for the cache depend on the update rates of all the users, update rates of the users affect each other indirectly. We also note that in Section \ref{sect:gen_soln}, where the caches are connected serially, we see this effect as a product of the terms, i.e., $\prod_{r=1}^{m}\frac{c_{r i}}{c_{r i}+\lambda_i}$. In this section, as the users are connected in parallel to a single cache, we see this effect as a summation of the terms, i.e., $\sum_{k=1}^{d} \frac{u_{ki}}{u_{ki}+\lambda_i}$.   

In the next section, we provide numerical results for the system with a single cache in Section \ref{sect:opt_soln}, with multiple caches in Section \ref{sect:gen_soln}, and with multiple users in Section~\ref{sect:mult_sol}.   

\section{Numerical Results} \label{sect:num_res}

In this section, we provide five numerical results. For these results, we use the following update arrival rates at the source
\begin{align} \label{lamb_dist}
    \lambda_i  = bq^i, \quad i = 1,\dots,n,
\end{align}
where $b> 0$ and $ 0< q\leq 1$ such that $\sum_{i=1}^{n}\lambda_i = a$. Note that with the update arrival rates at the source in (\ref{lamb_dist}), we have $\lambda_i\geq\lambda_j$ for $i\leq j$.  

\begin{figure}[t]
	\begin{center}
		\subfigure[]{%
			\includegraphics[width=0.49\linewidth]{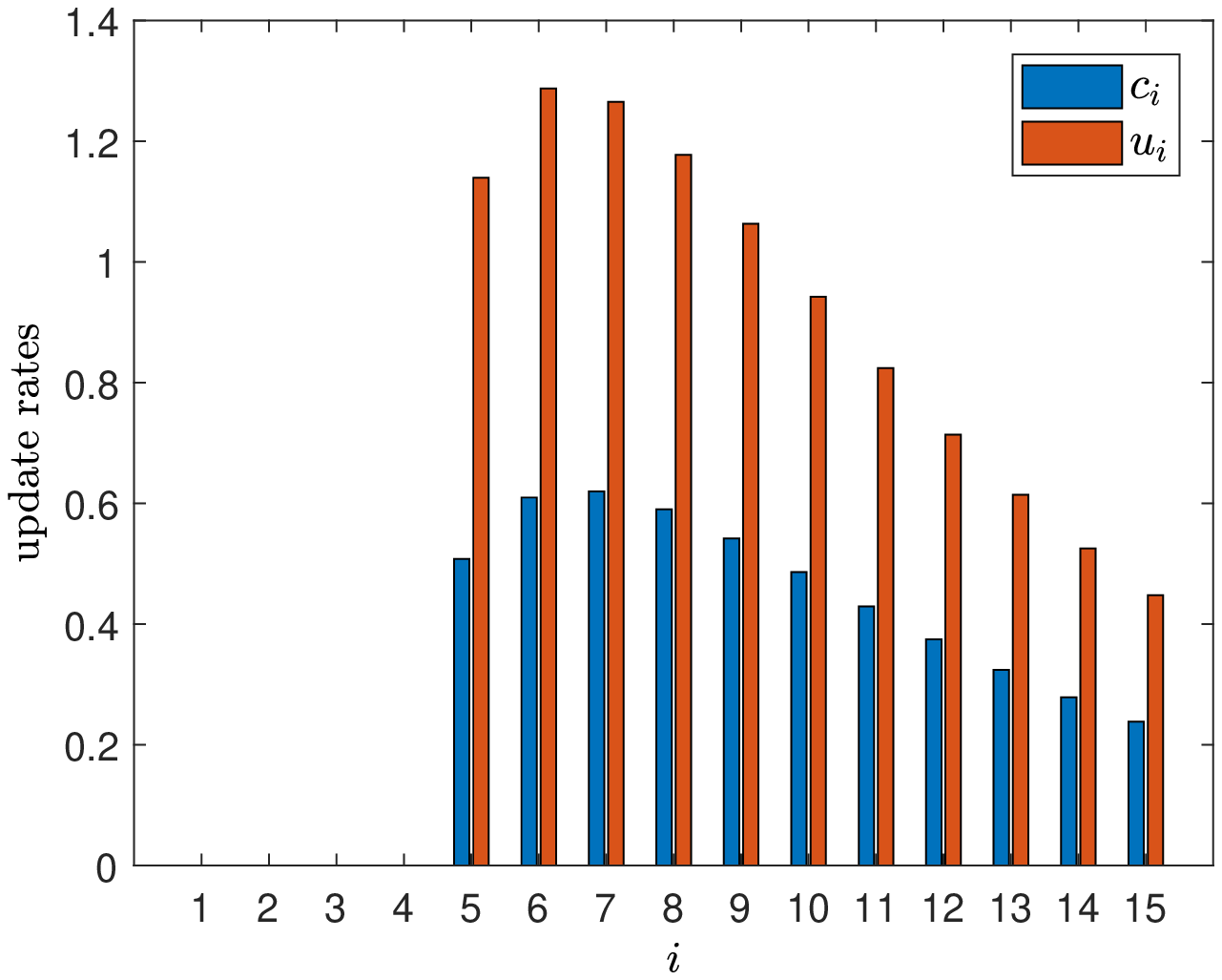}}
		\subfigure[]{%
			\includegraphics[width=0.49\linewidth]{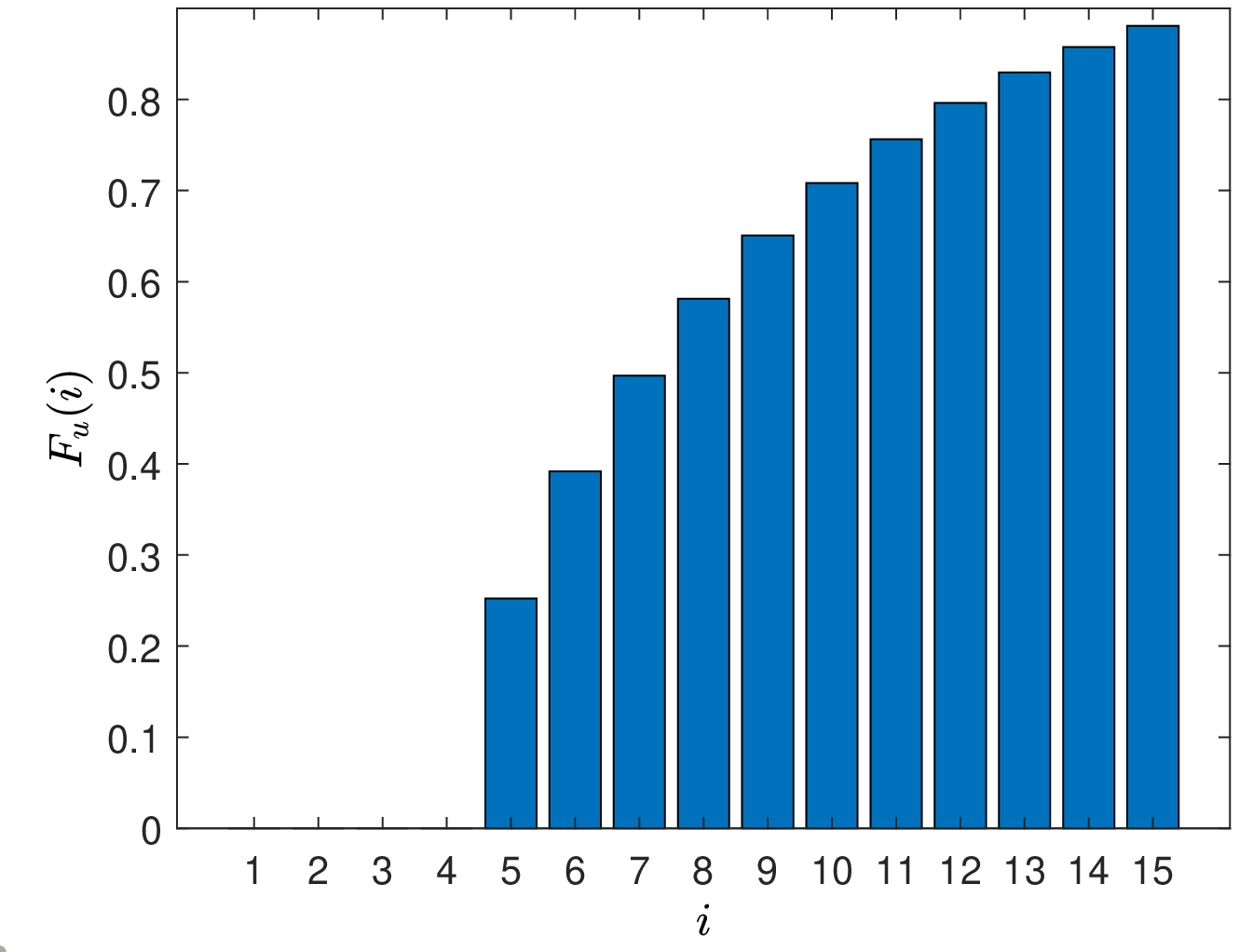}}
	\end{center}
	\caption{(a) Update rate allocation for the cache and the user for each file, and (b) the corresponding freshness $F_{u}(i)$, when $U=10$ and $C= 5$ with the file update rates at the source $\lambda_i$ given in (\ref{lamb_dist}), with $a =10$ and $q = 0.7$ for $n=15$.}
	\label{Fig:sim1}
\end{figure}

In example $1$, we take $a =10$, $q = 0.7$, and $n = 15$ in (\ref{lamb_dist}). For this example, we consider the system with a source, a single cache and a user. We choose the total update rate constraint for the cache as $C=5$, i.e., $\sum_{i=1}^{n}c_i\leq 5$, and for the user as $U = 10$, i.e., $\sum_{i=1}^{n}u_i\leq 10$. We initialize the file update rates at the user as $u_i = \frac{U}{n}$ for all $i$. We apply the alternating maximization method in Section \ref{sect:opt_soln} to find the update rates for the cache and for the user until the KKT conditions in (\ref{KKT1}) and (\ref{KKT2}) are satisfied. We see in Fig.~\ref{Fig:sim1}(a) that the first four files which are updated most frequently at the source are not updated by the cache and the user. As these files change too frequently at the source, their stored versions at the user become obsolete very quickly. In other words, updating files that change less frequently at the source brings more contribution to the overall information freshness at the user. The distributions of the update rates for the cache and for the user have similar trends as the update rates increase up to the seventh file for the cache and the sixth file for the user, and then update rates for the cache and the user decrease. Even though the update rates of the cache and the user for the slowly changing files are low, we see in Fig.~\ref{Fig:sim1}(b) that the freshness of the slowly varying files is higher compared to the rapidly changing files.                  

\begin{figure}[t]
	\begin{center}
		\subfigure[]{%
			\includegraphics[width=0.49\linewidth]{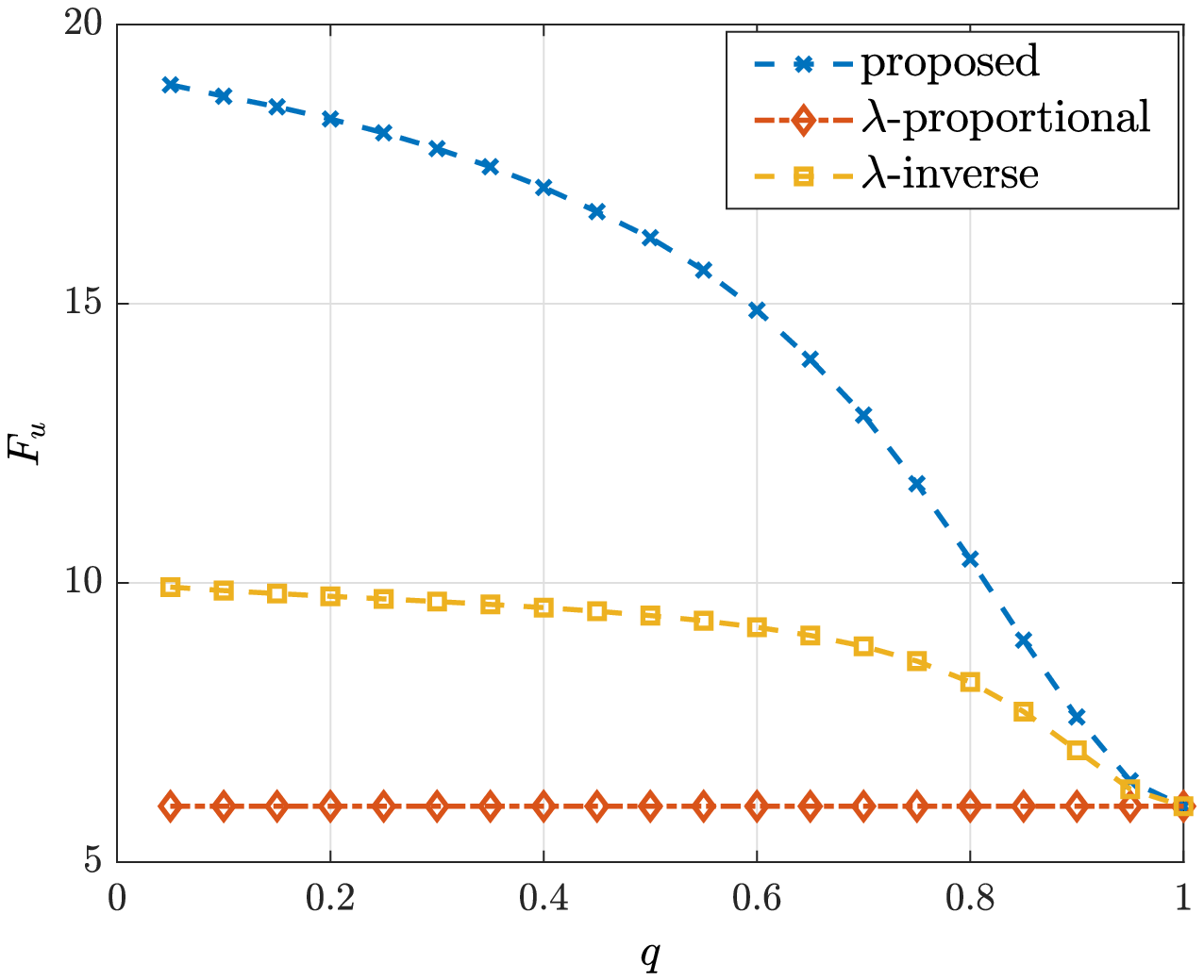}}
		\subfigure[]{%
			\includegraphics[width=0.49\linewidth]{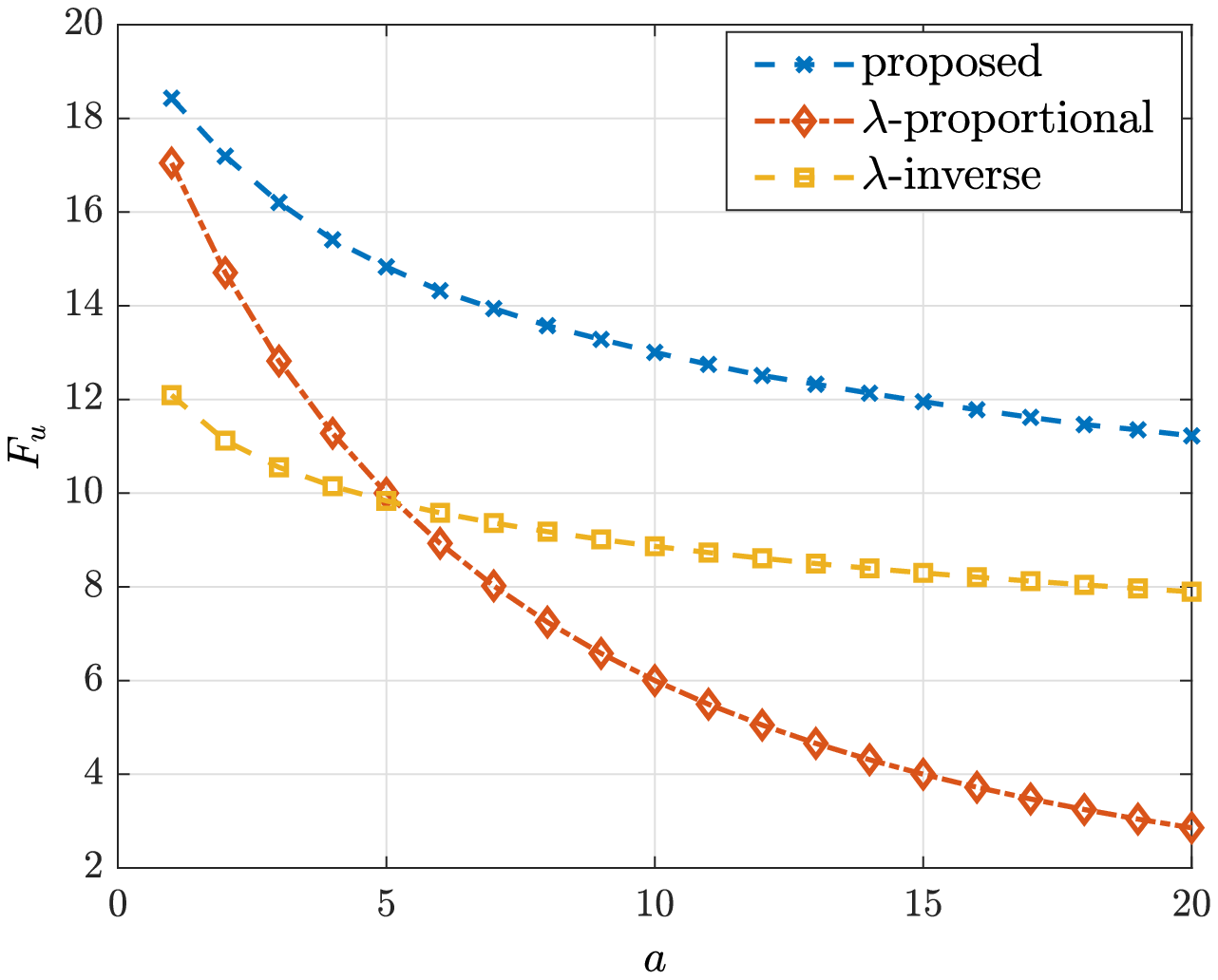}}
	\end{center}
	\caption{We compare the proposed update policy with the $\lambda$-proportional and the $\lambda$-inverse updating policies when $C=15$, $U=10$. We use $\lambda_i$ in (\ref{lamb_dist}) for $n=20$, (a) $a=10$, and $0<q\leq 1$ and (b) $q=0.7$ and $a=1,\dots,20$. }
	\label{Fig:sim5}
\end{figure}

 In example $2$, we consider the same system as in example $1$, and compare the performance of the proposed updating policy which maximizes the freshness at the user with two other baseline updating policies. As the first baseline policy, we consider $\lambda$-\textit{proportional} update policy where the update rates for the cache and for the user are chosen as $c_i =  \frac{\lambda_i C}{\sum_{i=1}^n \lambda_i}$ and $u_i =  \frac{\lambda_i U}{\sum_{i=1}^n \lambda_i}$ for all $i$, respectively. With this policy, the freshness of the $i$th file at the user is $F_u(i) = \frac{U}{U+a}\frac{C}{C+a}$. Thus, this update policy may be desirable if the user wants to have the same level of freshness for all files. As the second baseline policy, we consider $\lambda$-\textit{inverse} update policy where the update rates for the cache and for the user are given by $c_i = C \frac{\frac{1}{\lambda_i }}{\sum_{i=1}^n \frac{1}{\lambda_i}}$ and $u_i =  U\frac{\frac{1}{\lambda_i }}{\sum_{i=1}^n \frac{1}{\lambda_i}}$ for all $i$, respectively. The motivation for $\lambda$-inverse update policy stems from the fact that updating slowly varying files at the source with higher update rates brings more contribution to the overall freshness. In this example, we take $C=15$, $U=10$, and $\lambda_i$ in (\ref{lamb_dist}) with $n=20$. In Fig.~\ref{Fig:sim5}(a), we fix the total file update rate at the source, i.e., $a=10$, and vary the distribution of the file update rate at the source by choosing $0<q\leq 1$. We observe in Fig.~\ref{Fig:sim5}(a) that proposed update policy achieves the highest freshness compared to the considered baseline policies. We see that the $\lambda$-proportional policy achieves a constant level of freshness independent of the distribution of the file change rate at the source, but this policy gives the lowest freshness compared to others. We note that when $\lambda_i$s are evenly distributed, i.e., when $q$ is close to $1$, we see that all three updating policies achieve similar levels of freshness. In Fig.~\ref{Fig:sim5}(b), we fix $q=0.7$ and increase the file change rate at the source, i.e., $a=1,\dots,20$. We observe in Fig.~\ref{Fig:sim5}(b) that the proposed update policy achieves the highest freshness. When the total file change rate at the source is low, $\lambda$-proportional update policy achieves freshness close to the proposed update policy, whereas when the total file change rate at the source is high, $\lambda$-inverse achieves freshness close to the proposed update policy.

\begin{figure}[t]
	\centerline{\includegraphics[width=0.75\columnwidth]{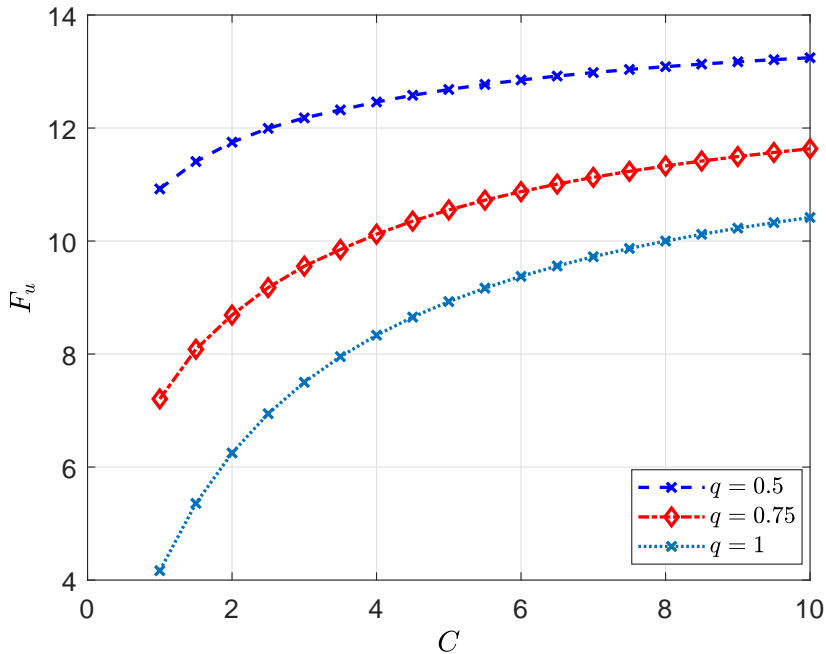}}
	\caption{Total freshness of the user $F_u$ with respect to $C$, when $\lambda_i$ are given in (\ref{lamb_dist}), with $a=2$ and $q= 0.5,0.75, 1$ for $n= 15$. }
	\label{Fig:sim2}
\end{figure}

In example $3$, we consider the same system as in examples $1$ and $2$ but this time we examine the effect of file update rates at the source over the information freshness at the user. We take $\lambda_i$ in (\ref{lamb_dist}) with $a=2$, $n=15$ for $q= 0.5, 0.75,1$. We note that a smaller $q$ corresponds to a less even (more polarized) distribution of file change rates at the source. We choose the total update rate for the user as $U = 10$ and vary the total update rate for the cache as $C = 1, 1.5, \dots, 10$. For each $C$ and $q$ values, we initialize $u_i = \frac{U}{n}$ for all $i$ and apply the alternating maximization method proposed in Section \ref{sect:opt_soln} until convergence. We see in Fig.~\ref{Fig:sim2} that when the distribution of the change rates of the files are more polarized, i.e., when $q$ is small, the overall information freshness at the user is larger as the freshness contribution from the slowly varying files can be utilized more with the more polarized distributions of file change rates at the source. We also note that for a fixed $\lambda_i$ distribution, the freshness of the user increases with the total update rate at the cache $C$ due to the fact that the user gets fresh files more frequently from the cache as the freshness of the files at the cache increases with $C$.            
\begin{figure}[t]
	\begin{center}
		\subfigure[]{%
			\includegraphics[width=0.49\linewidth]{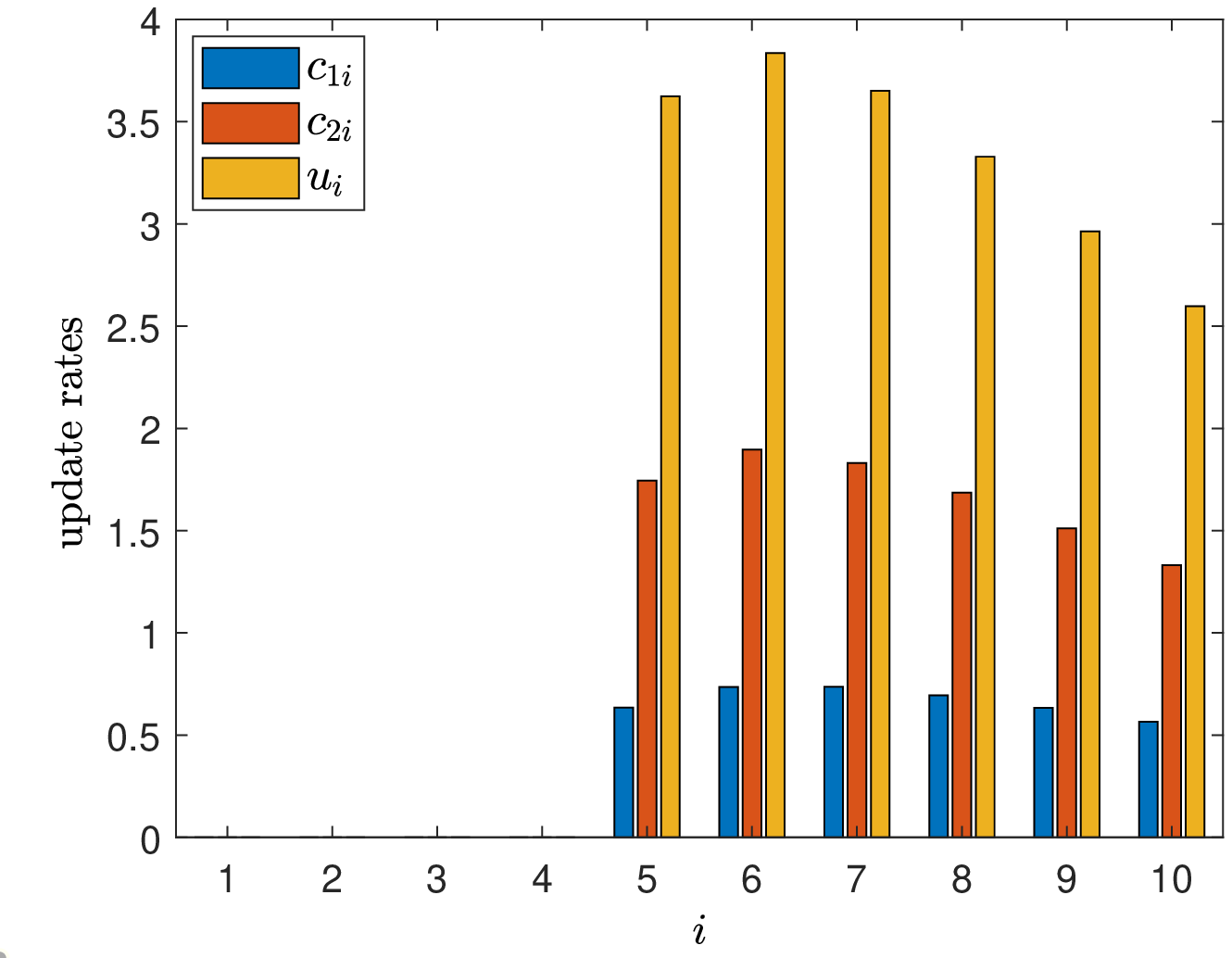}}
		\subfigure[]{%
			\includegraphics[width=0.49\linewidth]{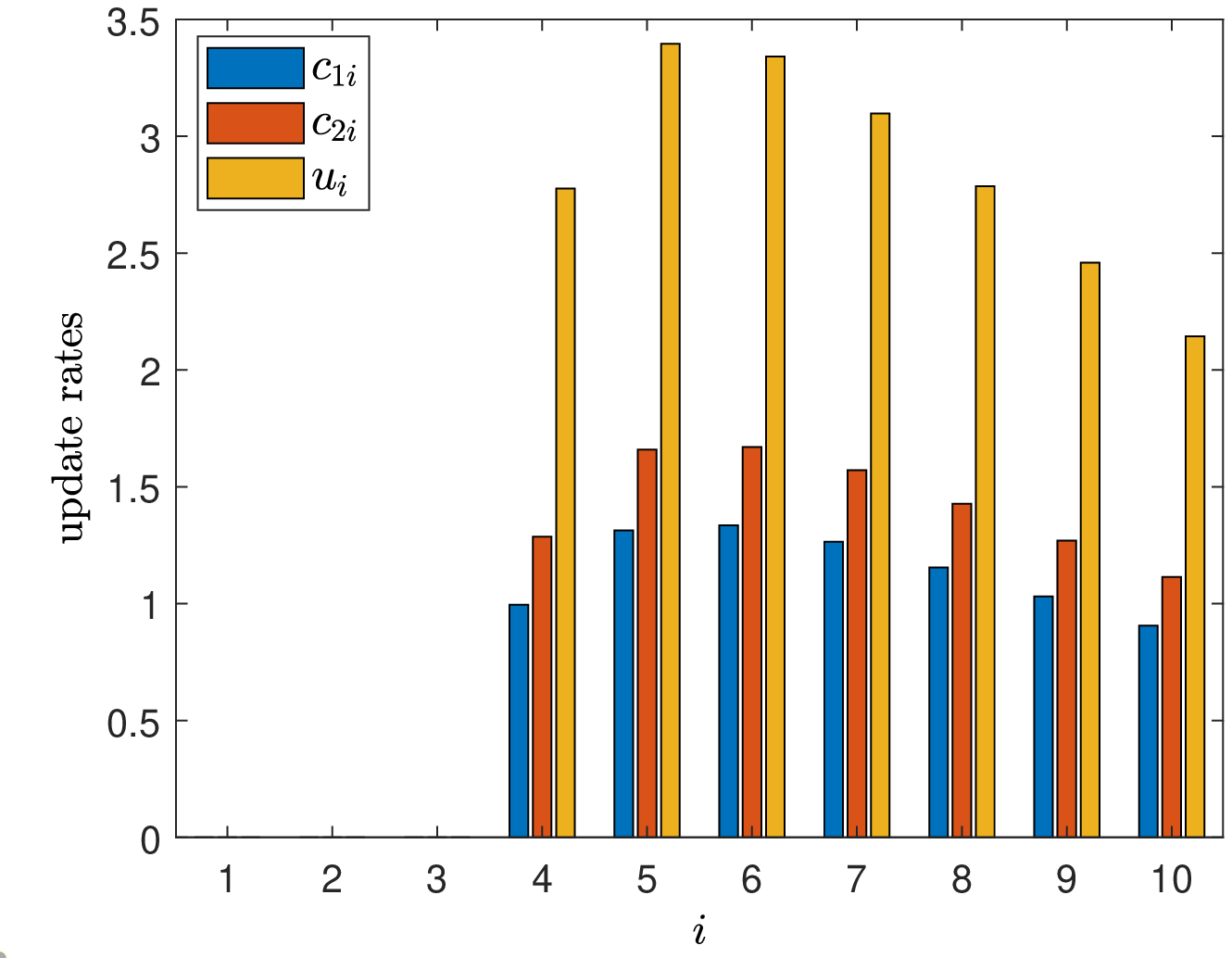}}\\
		\subfigure[]{%
			\includegraphics[width=0.49\linewidth]{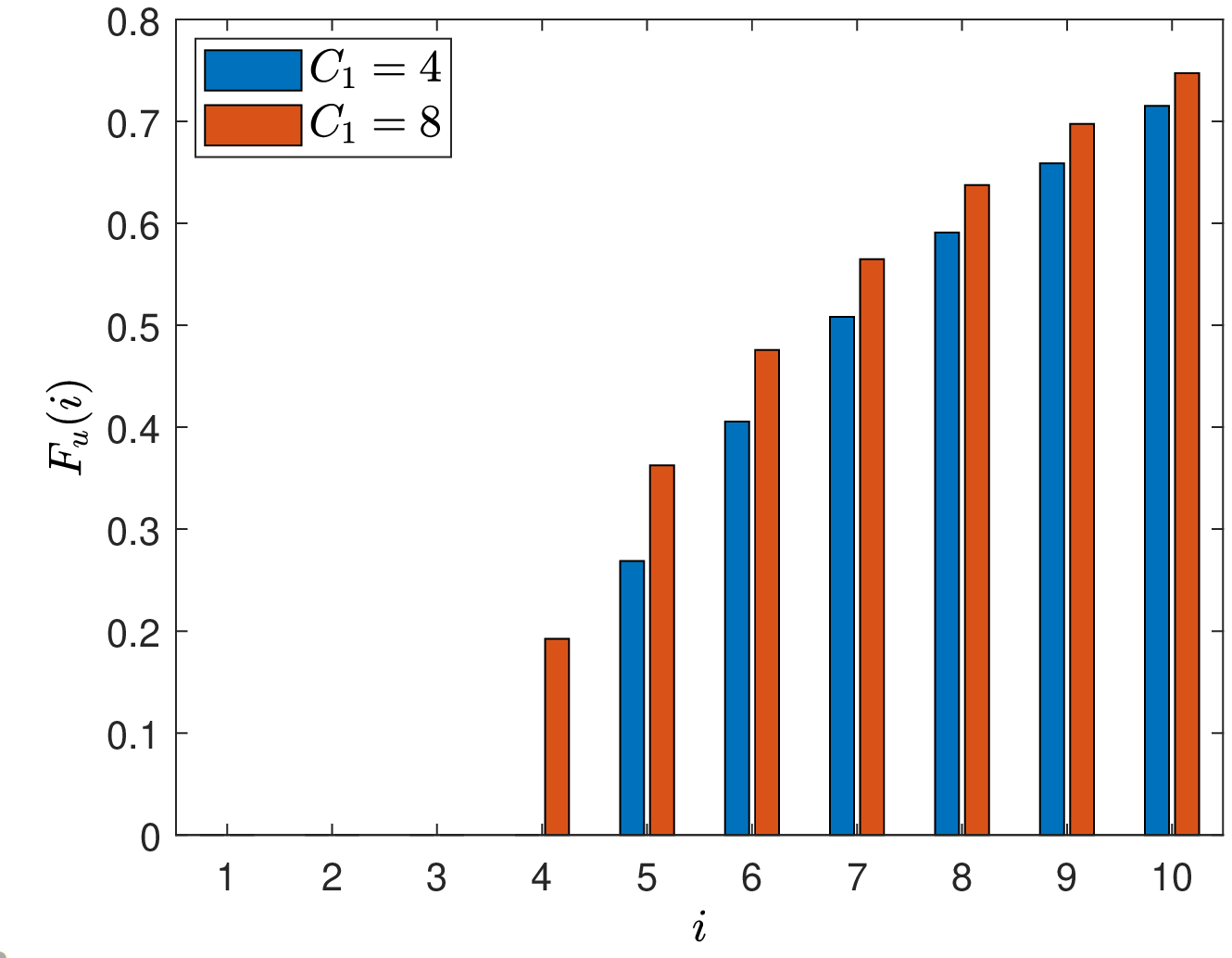}}
	\end{center}
	\caption{The update rates of the caches and the user when the total update rate constraint for the user is $U = 20$ and, for the second cache is $ C_2 = 10$. Total update rate constraint for the first cache is (a) $C_1 = 4$, and (b) $C_1 = 8$. The freshness of the files at the user is shown in (c).}
	\label{Fig:sim3}
\end{figure}

In example $4$, we consider a system where there are two caches placed in between the source and the user with $\lambda_i$ as given in (\ref{lamb_dist}) with $a = 10$, $q=0.7$, and $ n = 10$. We take the total update rate constraint for the user as $U=20$ and, for the second cache as $C_2 = 10$. We initialize the update rates at the user and at the second cache as $u_i = \frac{U}{n}$ and $c_{2i}= \frac{C_2}{n}$ for all $i$, respectively, and apply the alternating maximization method proposed in Section \ref{sect:gen_soln}, for the total update rate constraint for the first cache as $C_1 = 4,8$. We observe that the update rates for the caches and for the user have similar trends as shown in Fig.~\ref{Fig:sim3}(a) for $C_1 = 4$ and in Fig.~\ref{Fig:sim3}(b) for $C_1 = 8$. We note that the update rates of the caches in (\ref{soln_gen_c_i_2}) depend directly on the update rates of the other caches and also of the user. Similarly, the update rates of the user in (\ref{opt_rate_user_gen}) depend directly the update rates of all caches. That is why even though the total update rate constraints for the second cache and for the user remain the same in Fig.~\ref{Fig:sim3}(a)-(b), we see that the update rates at the second cache and at the user also change depending on the update rates at the first cache. In Fig.~\ref{Fig:sim3}(c), we observe that increasing the total update rate constraint for the first cache improves the freshness of every file except the first three files as the total update rate constraints for the caches and for the user are not high enough to update the most rapidly changing files. Furthermore, the improvement on the freshness of the rapidly changing files is more significant than the others as the freshness of the files at the user is a concave increasing function of $c_{1i}$.          

\begin{figure}[t]
	\begin{center}
		\subfigure[]{%
			\includegraphics[width=0.49\linewidth]{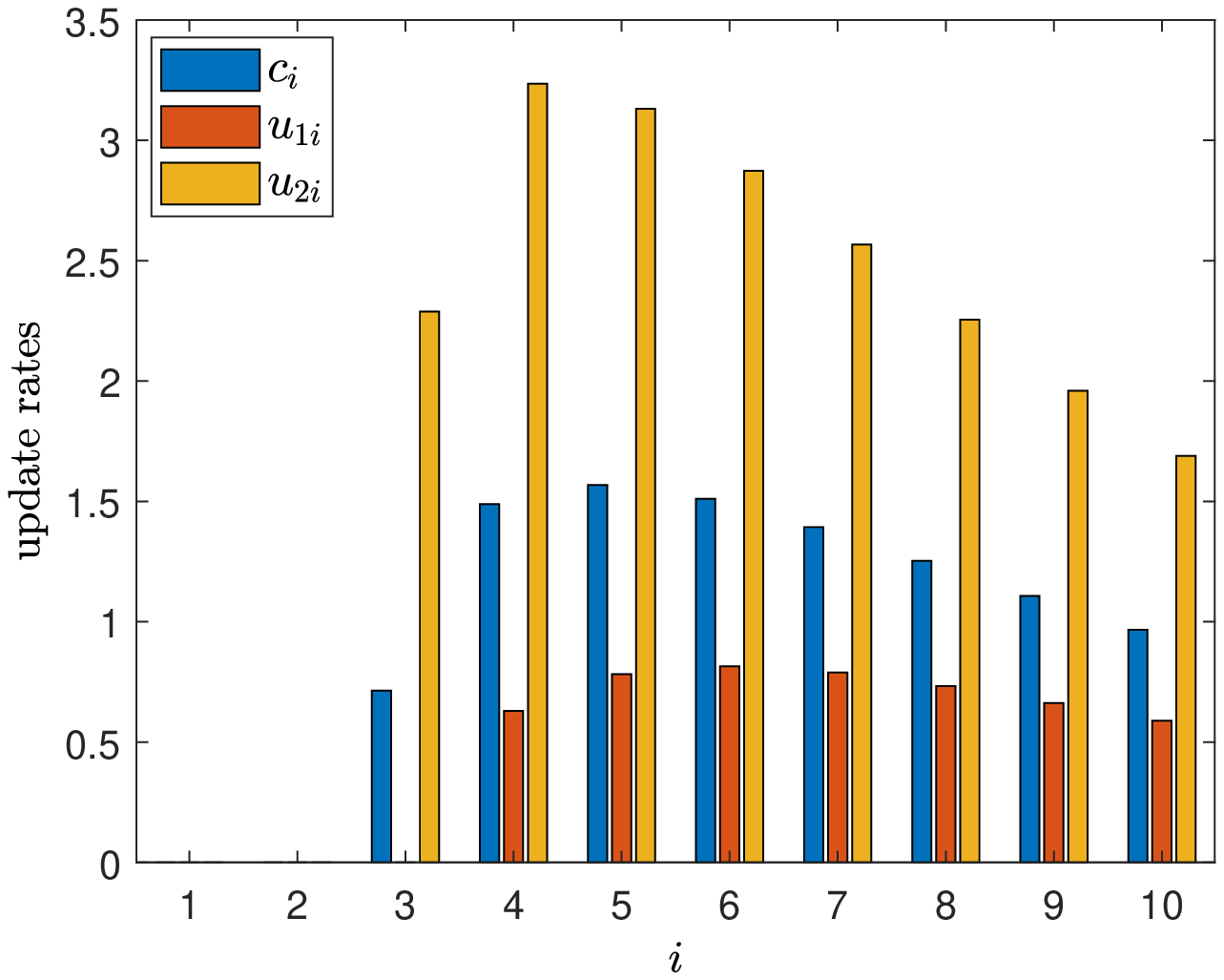}}
		\subfigure[]{%
			\includegraphics[width=0.49\linewidth]{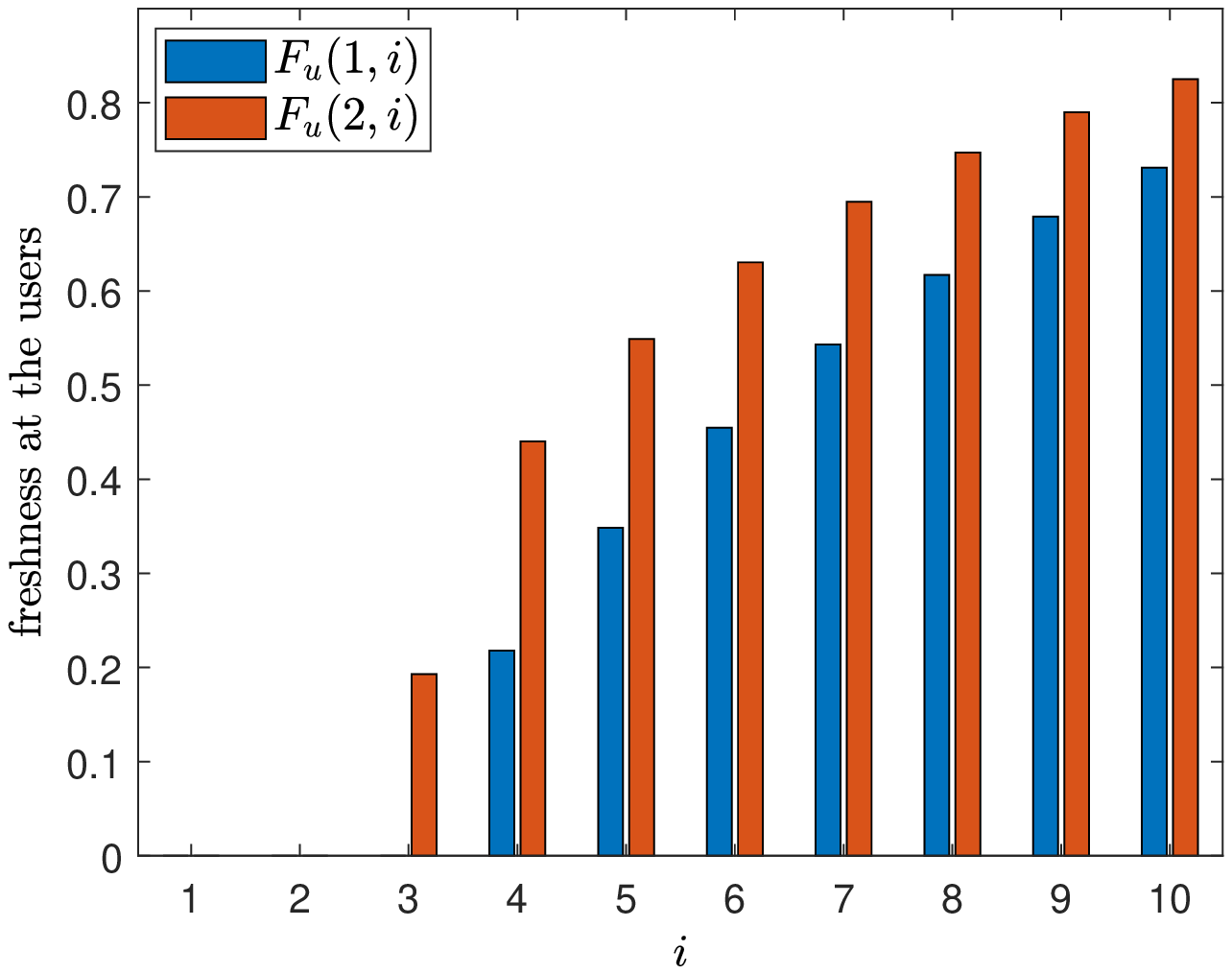}}
	\end{center}
	\caption{A system with a source, a single cache and two users: (a) The update rates of the cache and of the users, (b) the freshness of each file at the users.}
	\label{Fig:sim4}
\end{figure}

In example $5$, we consider the system where there is a source, a cache and two users connected to the cache. For this example, we use $\lambda_i$ in (\ref{lamb_dist}) with $a= 10$, $q=0.7$ and $n= 10$. We take the total update rate constraint for the cache as $C=10$, for the first user as $U_1 = 5$ and for the second user as $U_2 = 20$. We initialize the cache update rates as $c_i = \frac{C}{n}$ for all $i$, and apply the proposed alternating maximization based method to find $(c_i,u_{1i},u_{2i} )$ that satisfy the KKT conditions in  (\ref{KKT3_1}) and (\ref{KKT3_2}). The update rates of the cache and the users are shown in Fig.~\ref{Fig:sim4}(a) where we see that even though the update rate of the third file at the first user is equal to zero, the cache still updates the third file for the second user. We see in Fig.~\ref{Fig:sim4}(b) that the freshness of the files at the second user is higher as the total update rate of the second user is higher compared to the first user. The freshness difference between the users decreases for the slowly changing files at the source.

\section{Conclusion and Future Directions} \label{sect:dis}
In this paper, we first considered a cache updating system with a source, a single cache and a user. We found an analytical expression for the average freshness of the files at the cache and also at the user. We generalized this setting to the case where there are multiple caches placed in between the source and the user. Then, we provided an alternating maximization based method to find the update rates for the cache(s) and for the user to maximize the total freshness of the files at the user. We observed that for a given set of update rates of the user (resp. of the cache), the optimal rate allocation policy for the cache (resp. for the user) is a threshold policy where the frequently changing files at the source may not be updated by the user and also by the cache. Finally, we considered a system with multiple users refreshed via a single cache and found update rates for the cache and the users to maximize the overall freshness of the users.

As a future direction, one can consider a system where a user has access to multiple caches with different total update rate constraints. As a fresh file can be obtained through multiple caches, one can formulate a problem of finding the optimal rate allocation policies for the caches and for the user to maximize the freshness at the user. Another interesting direction to consider is the case where the caches can store only a portion of the files. In this case, one can consider the problem of joint content placement/assignment to the caches as well as the optimization of the update rates of each cache to maximize the information freshness for multiple users. While we considered only serially and parallel connected caches and users, one can investigate information freshness in arbitrarily connected networks.

\bibliographystyle{ieeetr}
\bibliography{IEEEabrv,lib_v1_melih}
\end{document}